\documentclass{article}
\usepackage[utf8]{inputenc}
\usepackage{stackengine}

\usepackage{geometry}
\usepackage[pdfencoding=auto, psdextra]{hyperref}
\usepackage{lineno}
\modulolinenumbers[5]

\usepackage{physics, siunitx, graphicx, amsmath, amssymb, mathtools, amsfonts, pgfplots, caption, subcaption, epstopdf, booktabs}
\usepackage[nocompress]{cite}
\usepackage[version=4]{mhchem}
\usepackage{comment}
\usepackage{pifont}

\usepackage{multirow}

\usepackage{tikz}
\usetikzlibrary{patterns,arrows}
\pgfplotsset{compat=1.13}

\newlength{\figwidth}
\newlength{\drop}
\numberwithin{equation}{section}

\usepackage{todonotes}
\newcommand{\jon}[1]{\todo[inline,caption={},color=orange!40]{Jon: #1}}

\newcommand{\ts}[1]{_{\text{#1}}}
\newcommand{\kin}[1]{\text{k} \in \{\text{#1}\}}

\newcommand{\Xav}[1]{\bar{#1}}
\newcommand{\YZav}[1]{\langle #1 \rangle}
\newcommand{\XYZav}[1]{\YZav{\Xav{#1}}}

\newcommand{\di}[1]{#1} 

\newcommand{\good}[1]{\textcolor{green}{#1}}
\newcommand{\ok}[1]{\textcolor{orange}{#1}}
\newcommand{\bad}[1]{\textcolor{red}{#1}}

\newcommand{\cmark}[1][green,fill=green]{\tikz[baseline=-0.5ex]\draw[#1,radius=3pt] (0,0) circle ;}%
\newcommand{\okmark}[1][orange,fill=orange]{\tikz[baseline=-0.5ex]\draw[#1,radius=3pt] (0,0) circle ;}%
\newcommand{\xmark}[1][red,fill=red]{\tikz[baseline=-0.5ex]\draw[#1,radius=3pt] (0,0) circle ;}%
\newcommand{\best}[1][green,fill=green]{\tikz[baseline=-0.5ex]\draw[#1,radius=3pt] (0,0) circle ;}%

\title{A Suite of Reduced-Order Models of a Single-Layer \\ Lithium-ion Pouch Cell}
\author{
    Scott G. Marquis
        \footnote{Mathematical Institute, Andrew Wiles Building, Woodstock Road, Oxford OX2 6GG, UK ({marquis@maths.ox.ac.uk}, {timms@maths.ox.ac.uk}, {please@maths.ox.ac.uk}, {chapman@maths.ox.ac.uk}).}
        \and
    Robert Timms \footnotemark[1] \footnote{The Faraday Institution, Quad One, Becquerel Avenue, Harwell Campus, Didcot, OX11 0RA, UK.} \and
    Valentin Sulzer \footnote{Department of Mechanical Engineering,
          Univ of Michigan, 2044 WE Lay Auto Lab, 1231 Beal Ave, Ann Arbor MI 48109-2133   ({vsulzer@umich.edu}).} \and
    Colin P. Please \footnotemark[1] \footnotemark[2] \and
    S. Jon Chapman  \footnotemark[1] \footnotemark[2]
}

\date{}

\usepackage{setspace}
\begin{document}

\maketitle

\begin{abstract}
For many practical applications, fully coupled three-dimensional models describing the behaviour of lithium-ion pouch cells are too computationally expensive. However, owing to the small aspect ratio of typical pouch cell designs, such models are well approximated by splitting the problem into a model for through-cell behaviour and a model for the transverse behaviour. In this paper, we combine different simplifications to through-cell and transverse models to develop a hierarchy of reduced-order pouch cell models. We give a critical numerical comparison of each of these models in both isothermal and thermal settings, and also study their performance on realistic drive cycle data. Finally, we make recommendations regarding model selection, taking into account the available computational resource and the quantities of interest in a particular study.

\end{abstract}

\section{Introduction}
Lithium-ion batteries are the most popular form of energy storage for many modern devices, with applications ranging from portable electronics to electric vehicles \cite{VanNoorden2014, scrosati2010, armand2008}. Improving both the performance and lifetime of these batteries by design changes that increase capacity, reduce losses and delay degradation effects is a key engineering challenge. Mathematical models allow possible improvements to be explored in an efficient manner before expensive and time consuming physical experiments are performed. However, a major hurdle is that existing battery models are computationally expensive, especially when seeking to resolve space- and time-dependent coupled electrical, thermal, and degradation effects within a battery. As a result, there is considerable interest in developing models that account for the key physical behaviours, but at a significantly reduced computational cost. In this paper, we derive a set of such simplified models, which are valid in various physically-relevant parameter regimes, by systematically reducing a detailed coupled electrochemical--thermal model of a lithium-ion pouch cell.

In \cite{sulzer_part_I}, we introduced a detailed fully-coupled 3D Doyle--Fuller--Newman (DFN) model of a lithium-ion pouch cell based upon the classical 1D DFN model developed by Newman and collaborators \cite{newman1962, newman2012, doyle1993}. A number of other similar higher-dimensional DFN-based models exist in the literature \cite{kosch2018, farag2017, baker1999, hosseinzadeh2018, kim2008, northrop2015}). This 3D DFN model allows for non-uniform behaviour in the current and temperature profiles in the transverse dimensions of the pouch. Capturing these non-uniform effects is of great importance for understanding how the cell may degrade in an uneven manner \cite{rieger2016, birkl2017degradation}, which in turn adversely affects cell performance and lifetime \cite{kosch2018}. However, the full 3D DFN model is very computationally expensive, motivating us to develop simpler models which still retain the essential physics. By exploiting the large geometric aspect ratio of the cell, and the large ratio of the thermal voltage to the typical Ohmic drop in the current collectors, we derived the simplified ``2+1D'' DFN model. The 2+1D DFN comprises a collection of 1D DFN models describing the through-cell electrochemistry, coupled via a 2D electrical problem in the current collectors and a 2D thermal problem. Similar 2+1D models, also referred to as ``potential pair'' models, have been exploited in an ad-hoc way in many battery simulations. Within the 2+1D structure, various models of have been employed to represent the 1D through-cell electrochemistry, these include: the 1D DFN \cite{lee2013}; a nonlinear resistor fitted to an electrochemical model \cite{gerver2011}; and data \cite{kim2008}. In this paper, we begin from the 2+1D DFN model and, via asymptotic methods, systematically derive a suite of computationally simpler models.

\begin{figure}[htbp]
  \begin{center}
    \footnotesize
    \stackunder[\drop]{\includegraphics[width=\figwidth]{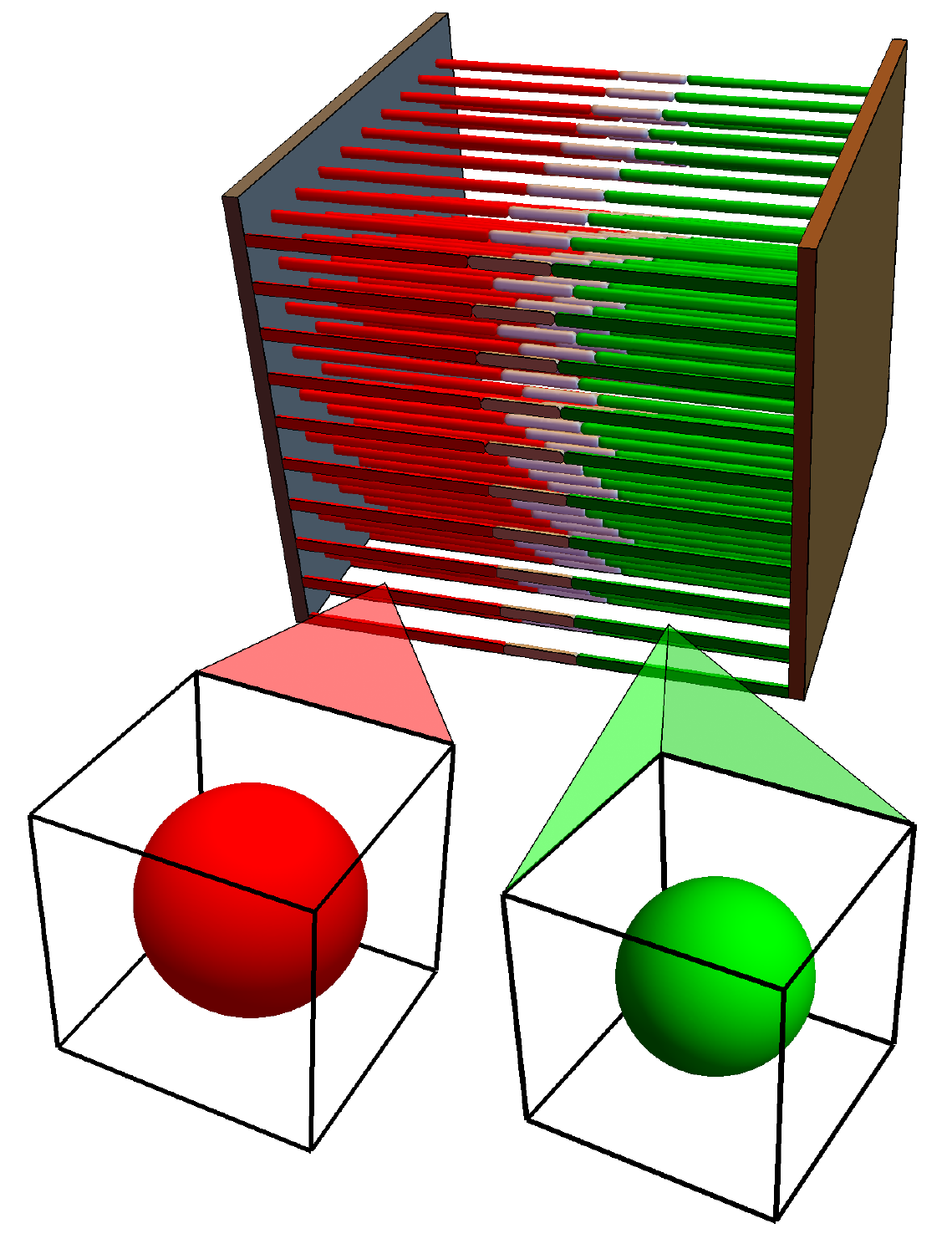}}{2+1D DFN (P4D)}
    \qquad
    \stackunder[\drop]{\includegraphics[width=\figwidth]{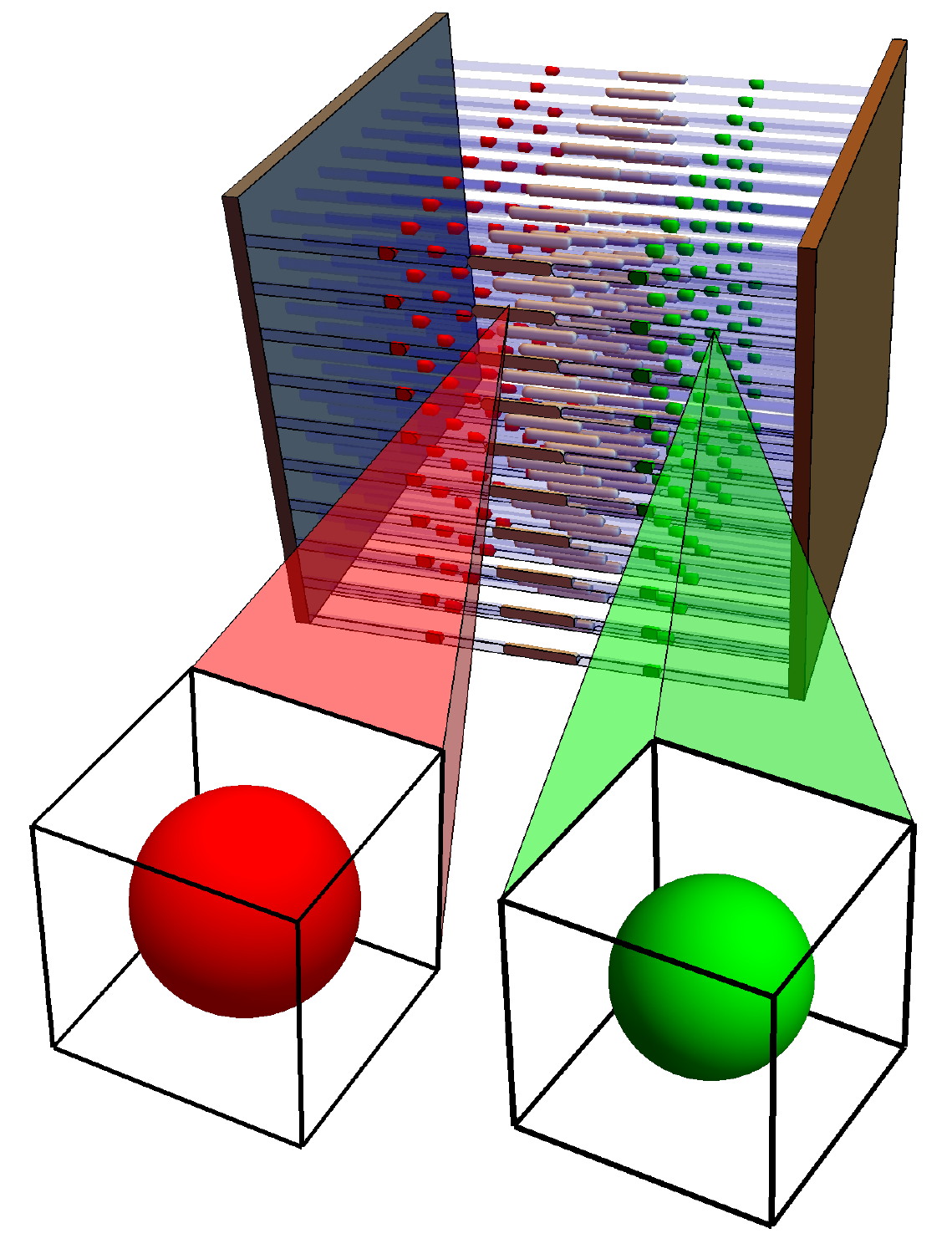}}{2+1D SPMe (P3D)}
    \qquad
    \stackunder[\drop]{\includegraphics[width=\figwidth]{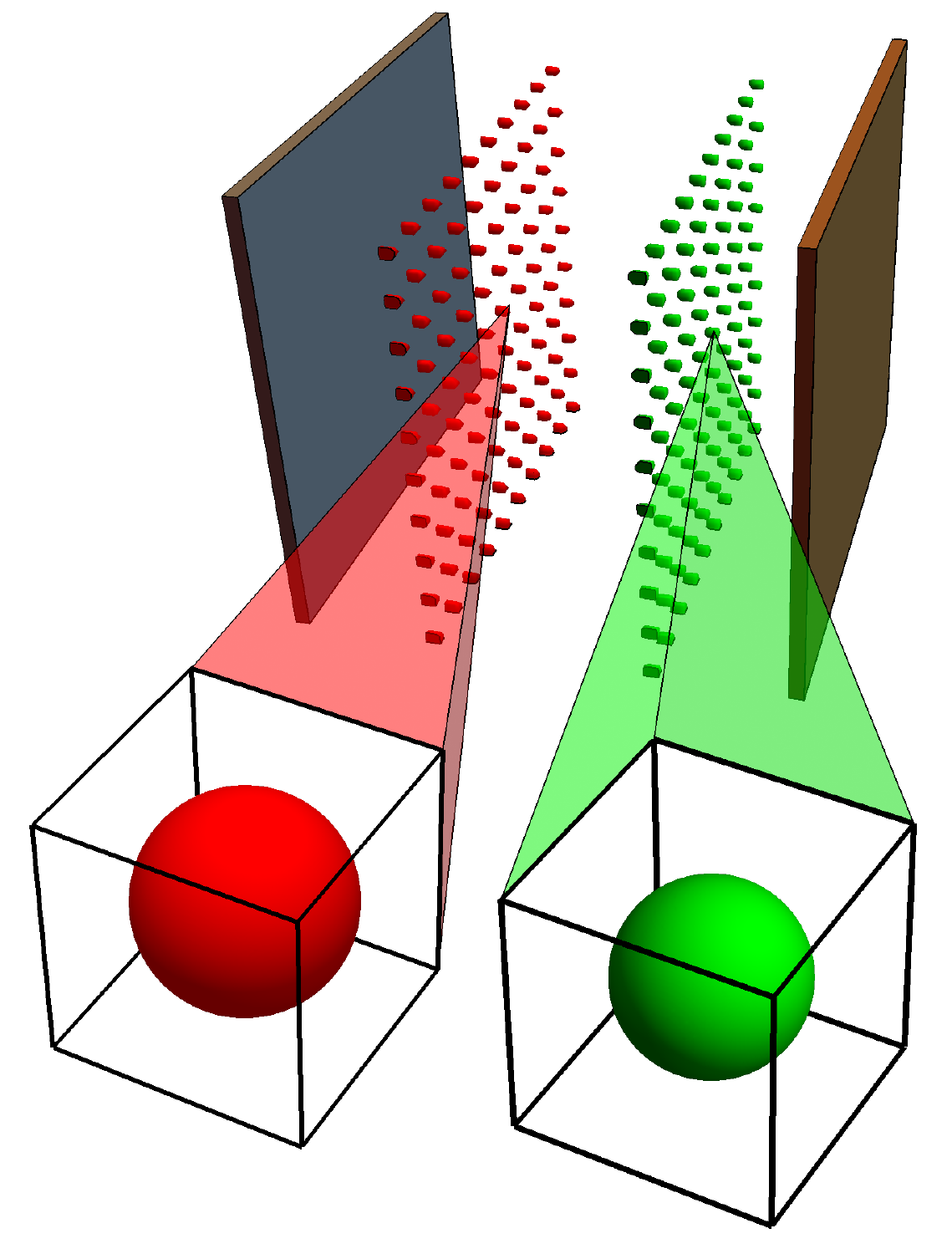}}{2+1D SPM (P3D)}
    \\[4mm]
    \stackunder[\drop]{\includegraphics[width=\figwidth]{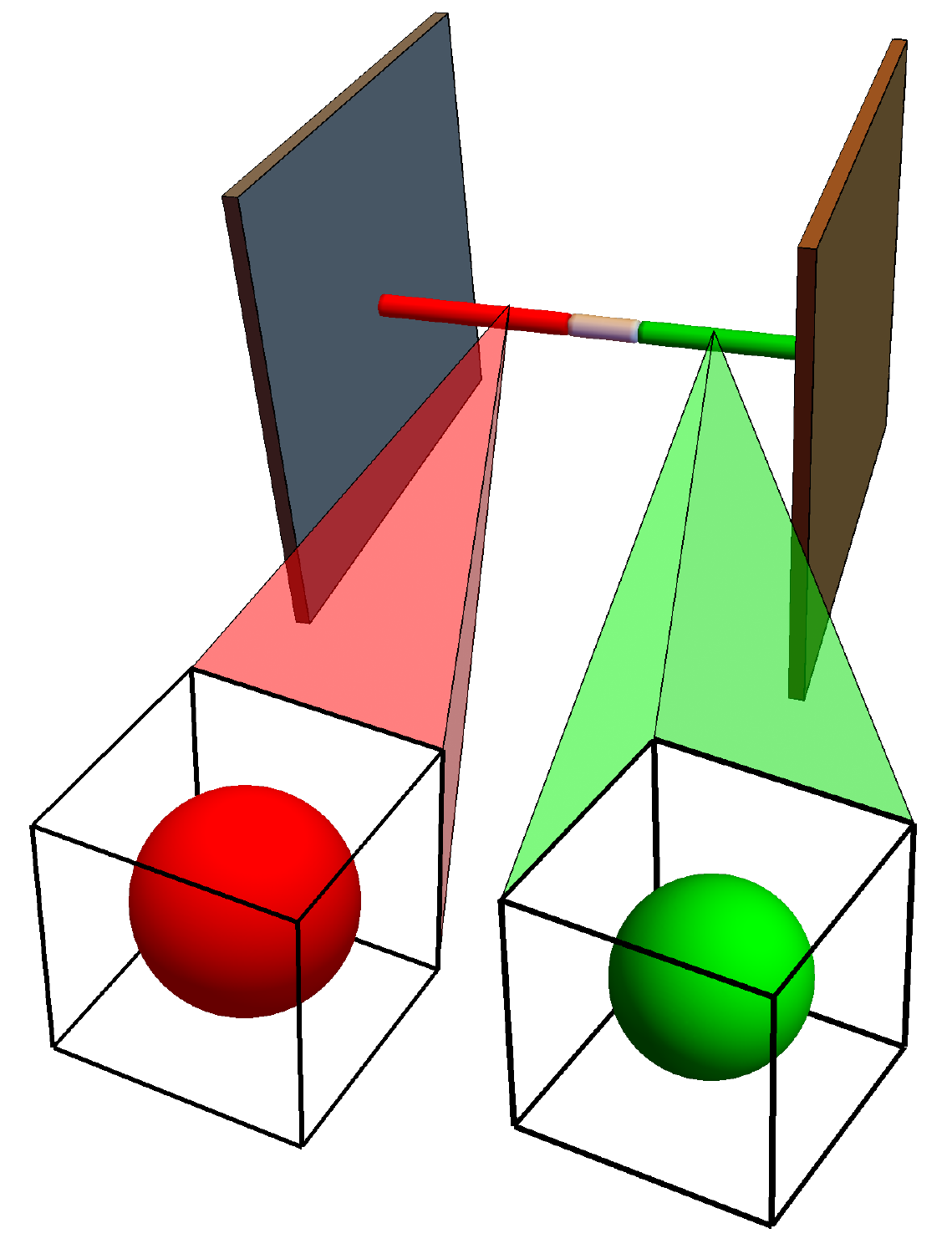}}{DFNCC (P2D)}
    \qquad
    \stackunder[\drop]{\includegraphics[width=\figwidth]{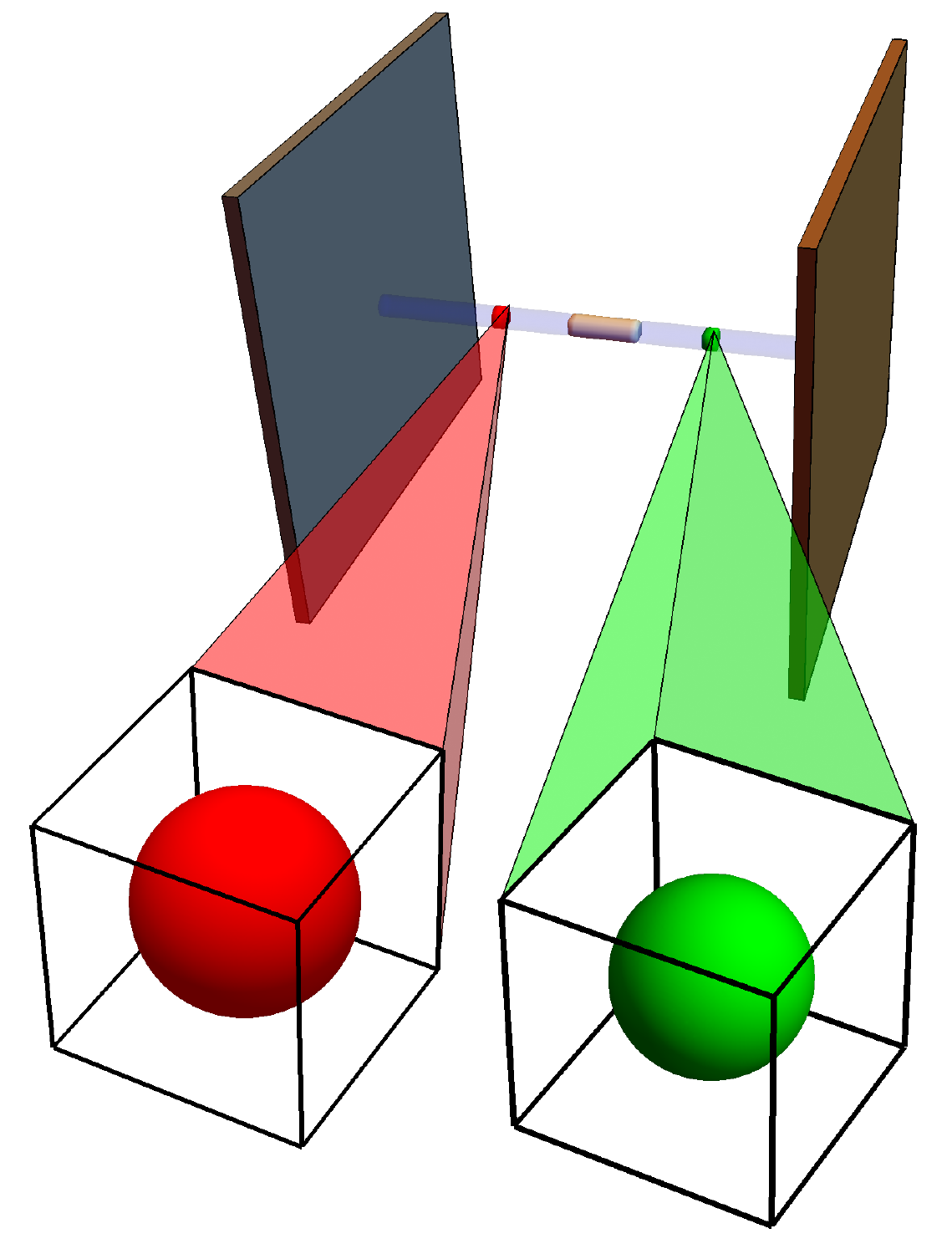}}{SPMeCC (P1D)}
    \qquad
    \stackunder[\drop]{\includegraphics[width=\figwidth]{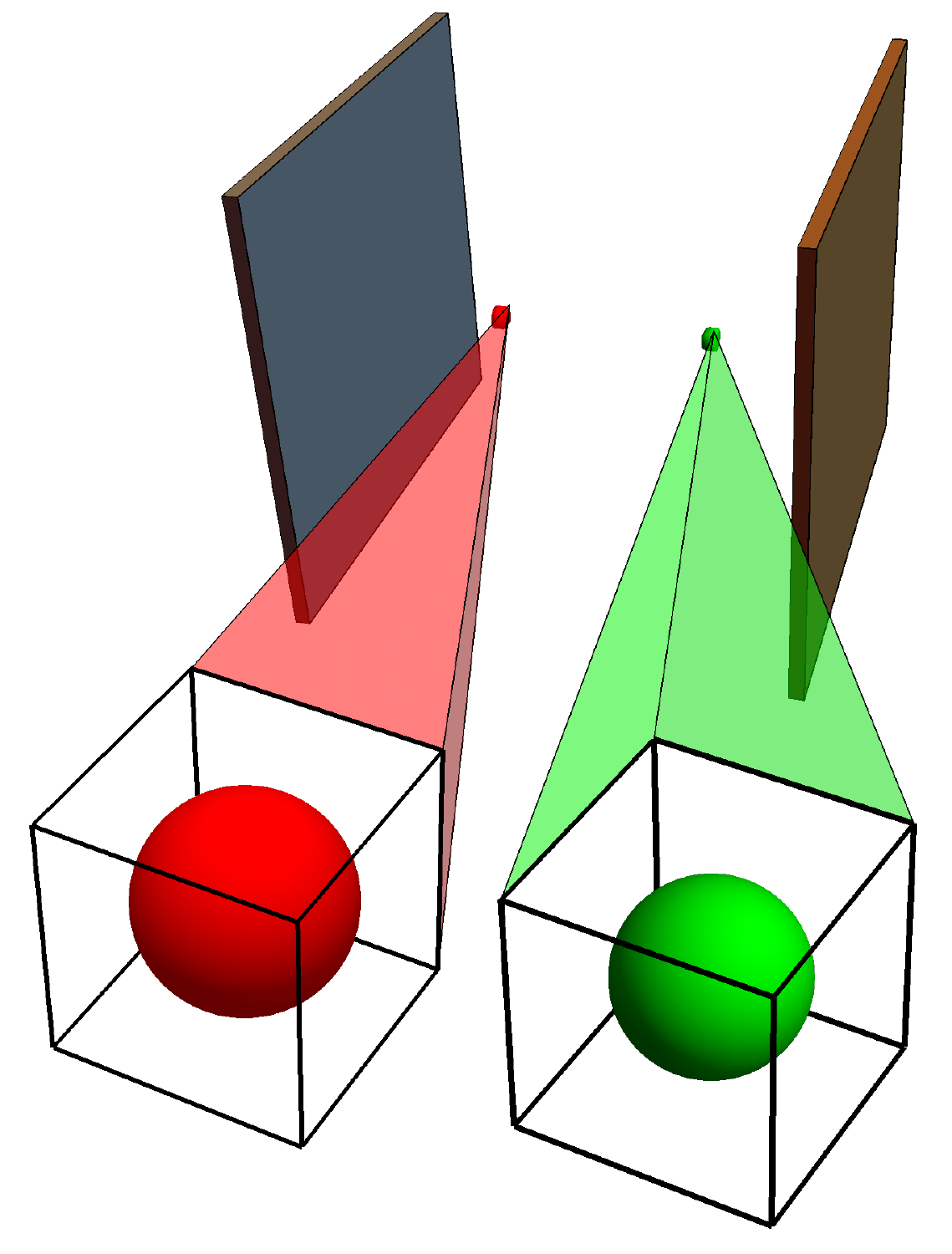}}{SPMCC (P1D)}
    \\[4mm]
    \stackunder[\drop]{\includegraphics[width=\figwidth]{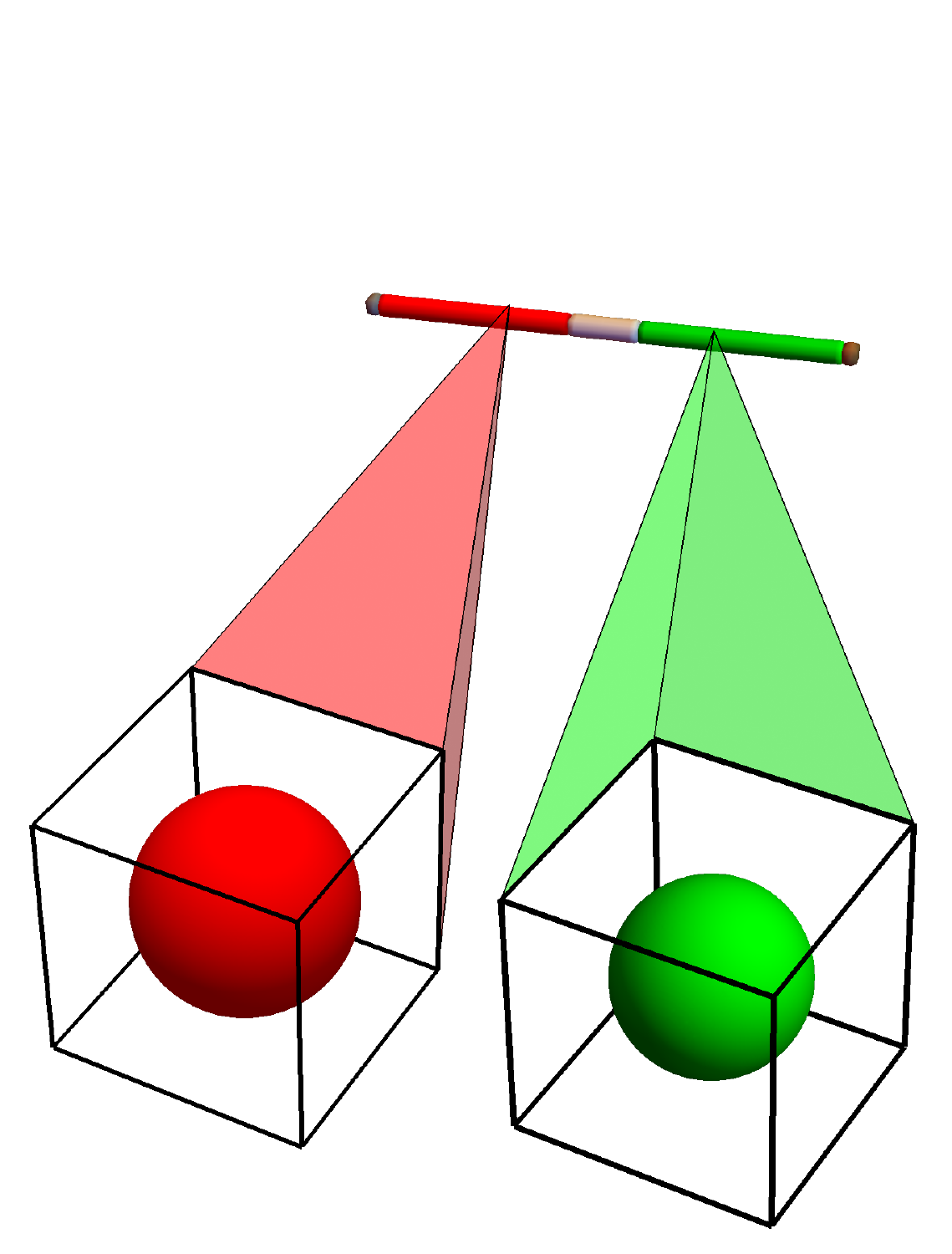}}{DFN (P2D)}
    \qquad
    \stackunder[\drop]{\includegraphics[width=\figwidth]{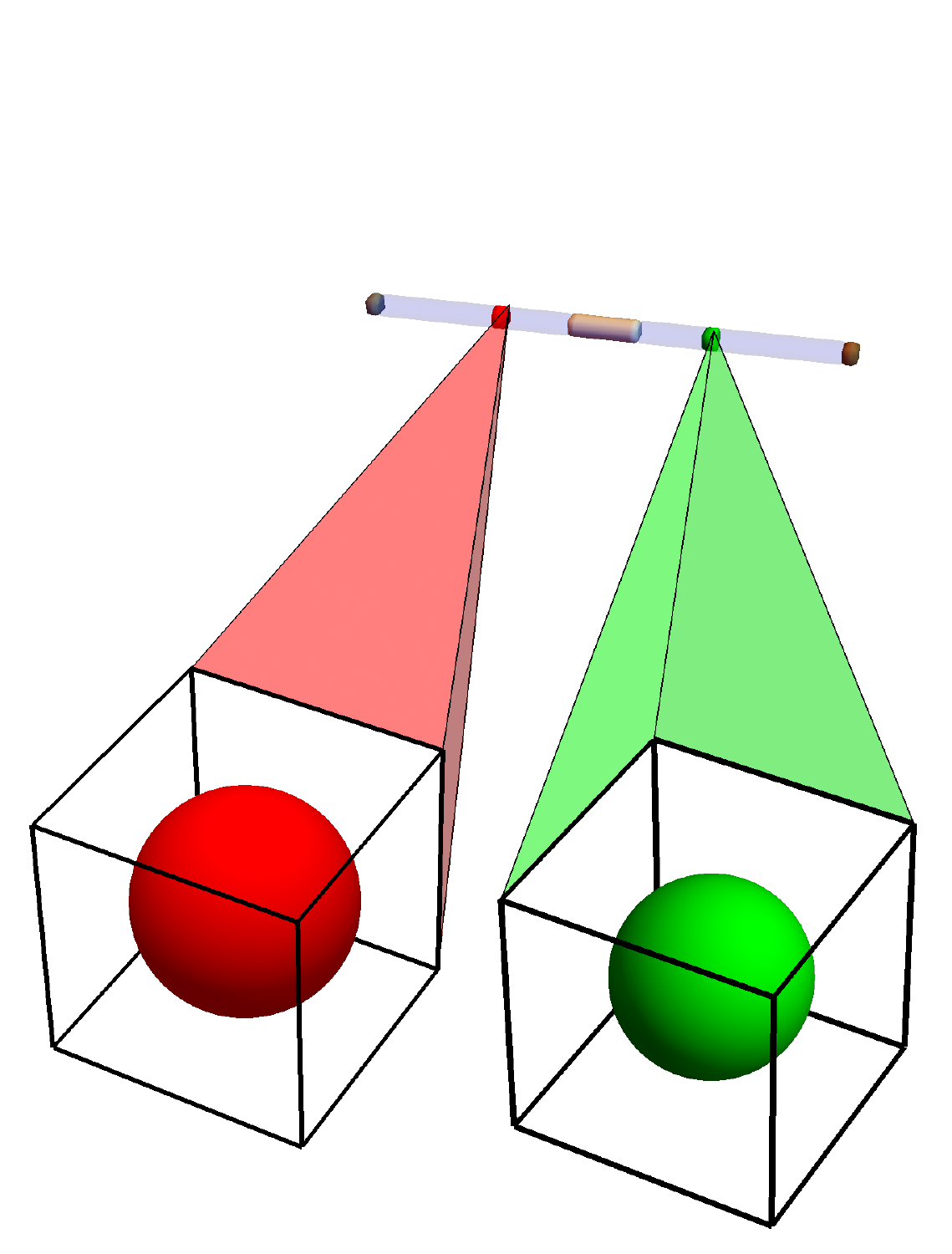}}{SPMe (P1D)}
    \qquad
    \stackunder[\drop]{\includegraphics[width=\figwidth]{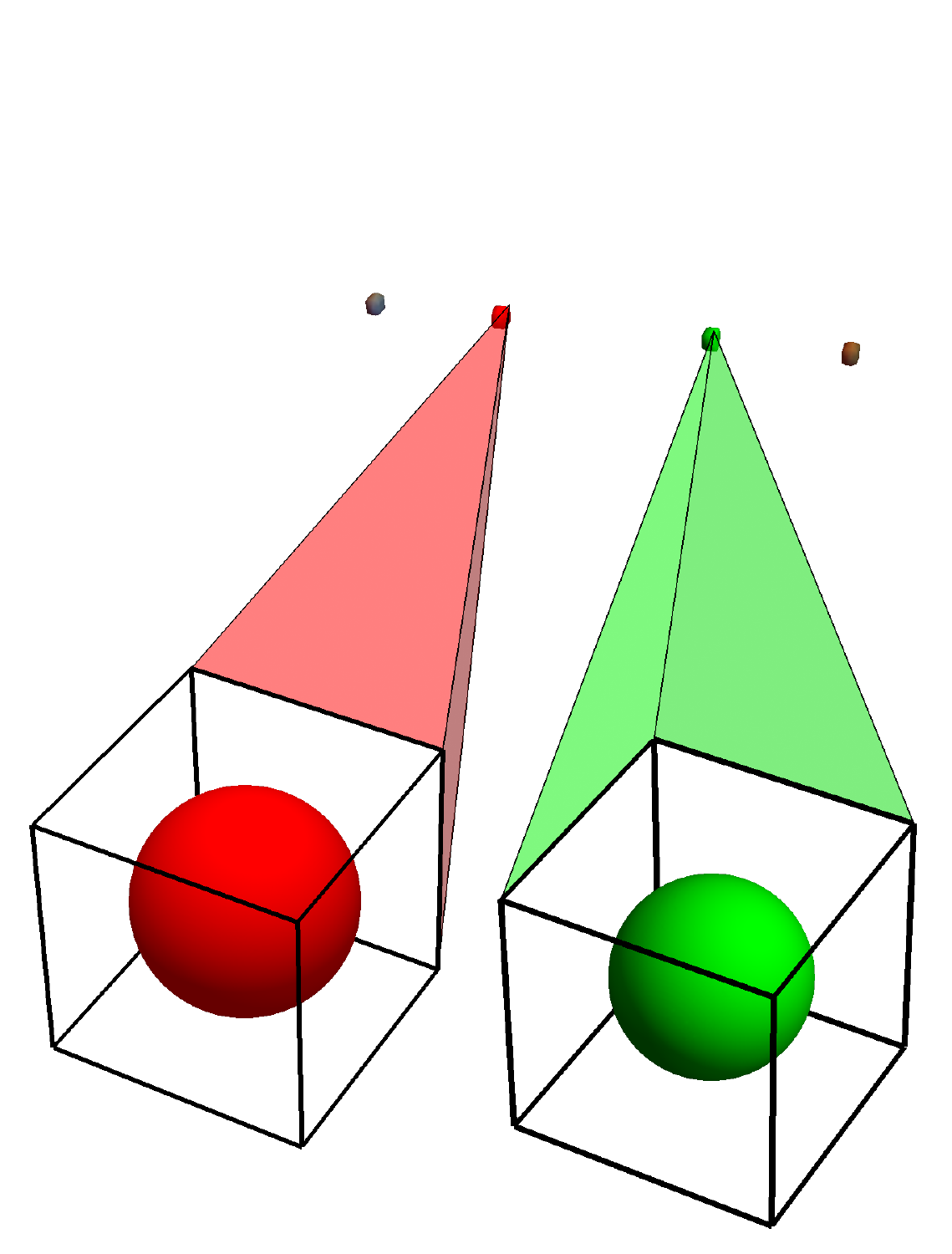}}{SPM (P1D)}
  \end{center}
  \caption{Schematic diagram of the models we consider, and their complexity.  DFN corresponds to  one-dimensional problem through the cell, at each point of which there is a radial problem for the concentration of Li in the active material. Thus, such a model is pseudo-two-dimensional (P2D). In 2+1 DFN, at every point of the current collectors a one-dimensional DFN model is solved, leading to a pseudo-four-dimensional (P4D) model. In the DFNCC, a single DFN model is solved, along with an uncoupled two-dimensional problem in the current collectors. SPM considers a single active particle in each electrode, leading to a one-dimensional model (P1D). In the SPMe, an extra one-dimensional equation for the electrolyte is added; such a model is still P1D.}
  \label{fig:model_schematic}
\end{figure}

The models that we compare in this paper are displayed pictorially in Figure~\ref{fig:model_schematic}. The 2+1D DFN, the most complex model we consider, is shown in the top-left panel. This model consists of a two-dimensional problem for the potentials in each current collector and a two-dimensional thermal problem, coupled to a one-dimensional DFN model at each point representing the through-cell electrochemistry. Proceeding downwards in Figure~\ref{fig:model_schematic}, represents making simplifications to the transverse model (i.e. the current collectors). The first simplification (the middle row) gives rise to a set of models we label `CC'. These models consist of a single `average' through-cell electrochemical model and a second decoupled two-dimensional problem for the current collector resistances (hence the name `CC') which can be solved `offline' before solving the through-cell model. On the bottom row of Figure~\ref{fig:model_schematic}, the effects of the current collectors are neglected entirely, with only a single representative through-cell electrochemical model solved. We refer to these as 0D transverse models. The model equations associated with each of these three transverse models are presented in \S\ref{sec:transverse-models}. 
Moving to the right in Figure~\ref{fig:model_schematic} represents making simplifications to the through-cell electrochemical model. The three through cell models that we consider are the DFN model, the single particle model with electrolyte (SPMe), and the single particle model (SPM). Formal derivations of the SPMe and SPM from the DFN can be found in \cite{Marquis2019}. The DFN model comprises one-dimensional equations for the electrode potentials, the electrolyte potentials, and the electrolyte concentrations that are posed on the macroscale (the thickness of the electrode). At each point in the model there is a radial problem posed on the microscale (the radius of an active particle) for the lithium concentrations in the particles. The DFN model is often referred to as ``pseudo two-dimensional'' (P2D) since at each point in the one-dimensional macroscale there is an additional one-dimensional microscale problem (i.e there is a particle at each point in the electrode). The SPMe replaces the collection of problems for the particles with a single average particle in each electrode. The equations for the average particles are coupled to a one-dimensional equation for the electrolyte concentration, and the electrode and electrolyte potentials are then recovered via algebraic expressions. Finally, the SPM discards the problem in the electrolyte, and consists of only a single representative particle in each electrode. The model equations for each of these through-cell electrochemical models are presented in \S\ref{sec:through-cell-models}. 
By combining any transverse model with any through-cell electrochemical model, we arrive at one of the nine models in Figure~\ref{fig:model_schematic}. For example, by choosing the `CC' transverse model and the SPMe for the through-cell model, we arrive at the SPMeCC model, which consists of an `average' SPMe model and decoupled current collector problem. Alternatively, by choosing a 0D transverse model and the SPM through-cell model, we arrive at the standard 1D SPM model. We shall compare each of the nine models in both the isothermal and thermal cases, and in particular assess the predictions of key output variables from each model (e.g. the terminal voltage, average temperature, etc.). By doing this, we aim to aid the reader in choosing the appropriate model for a particular application with given requirements of computational speed and accuracy. Our results are summarized in \S\ref{sec:comparison-summary} and we particularly guide the reader's attention to Table~\ref{tab:summary-table} for an overview of the strengths and weaknesses of each model. An implementation of these models is also available in the open-source battery modelling software PyBaMM \cite{pybamm}. 

\subsection{Notation}
Before stating the governing equations we comment on our notation. We denote electric potentials by $\phi$, current densities by $\boldsymbol{i}$, lithium concentrations\footnote{In the electrolyte $c$ denotes the lithium-ion concentrations.} by $c$, molar fluxes by $\boldsymbol{N}$, and temperatures by $T$. To distinguish potential, fluxes and concentrations in the electrolyte from those in the solid phase of the electrode, we use a subscript $\text{e}$ for electrolyte variables and a subscript $\text{s}$ for solid phase variables. To indicate the region within which each variable is defined, we include an additional subscript k, which takes one of the following values: n (negative electrode), p (positive electrode), cn (negative current collector), cp (positive current collector), or s (separator).
For example, the notation $\phi\ts{s,n}$ refers to the electric potential in the solid phase of the negative electrode. When stating the governing equations, we take the region in which an equation holds to be implicitly defined by the subscript of the variables. These regions are given by
\begin{align*}
    \Omega\ts{cn} &=  [-L\ts{cn},0]\times\Omega, &
    \Omega\ts{n} &=  [0,L\ts{n}]\times\Omega, &
    \Omega\ts{s} &=  [L\ts{n},L_x-L\ts{p}]\times\Omega, \\
    \Omega\ts{p} &=  [L_x-L\ts{p},L_x]\times\Omega,& 
    \Omega\ts{cp} &=  [L_x, L_x+L\ts{cp}]\times\Omega,
    \end{align*}
corresponding to the negative current collector, negative electrode, separator, positive electrode, and positive current collector, respectively,
where $\Omega = [0,L_y]\times[0,L_z]$
is the projection of the cell onto the $(y,z)$-plane. Here $L\ts{k}$ ($\kin{cn, n, s, p, cp}$) are the thicknesses of each component of the cell, $L_x = L\ts{n}+L\ts{s}+L\ts{p}$ is the distance between the two current collectors, and $L_y$ and $L_z$ are the cell width and height, respectively.

 We also introduce the notation $\partial\Omega\ts{tab,k}$ to refer to the negative and positive tabs ($\kin{cn, cp}$), $\partial\Omega\ts{ext,k}$ to refer to the external boundaries of region $\kin{cn, n, s, p, cp}$, and $\partial\Omega_{\text{k}_1, \text{k}_2}$ to refer to the interface between regions $\text{k}_1$ and $\text{k}_2$. For instance, the notation $\partial\Omega\ts{n,s}$ refers to the interface between the negative electrode and the separator. Finally, for $\kin{cn, cp}$  we use $\partial\Omega\ts{tab,k,$\perp$}$ to denote the projection of the tabs onto the $(y,z)$-plane, and
$\partial\Omega\ts{ext,k,$\perp$} =  \partial\Omega\setminus\partial\Omega\ts{tab,k,$\perp$}$ to denote the 
non-tab region of the boundary of the projection.

Derivatives in $x$ and $r$ are written out explicitly, and we define the gradient in the transverse direction as
\begin{equation}
  \nabla_\perp \equiv \pdv{y}\boldsymbol{e}_2 + \pdv{z}\boldsymbol{e}_3.
  \label{transversegrad}
\end{equation}
We use an overbar to denote an average in the $x$-direction, so that in each electrode
\begin{align}
    \Xav{f}\ts{n} := \frac{1}{L\ts{n}}\int_{0}^{L\ts{n}} f\ts{n} \, \dd{x}, \quad
    \Xav{f}\ts{p} := \frac{1}{L\ts{p}}\int_{L_x-L\ts{p}}^{L_x} f\ts{p} \, \dd{x}.
\end{align}
We use angled brackets to denote an average in the ($y,z$)-directions,
\begin{equation}
    \YZav{f} = \frac{1}{L_y L_z} \int_{\Omega} f \, \dd{y}\, \dd{z}.
\end{equation}
Quantities averaged over both the $x$-direction of an electrode and the $y$-$z$ directions are then denoted by $\YZav{\Xav{f\ts{k}}}$.

\section{Transverse Models}\label{sec:transverse-models}

Our starting point, the ``2+1D DFN model'', is a set of through-cell one-dimensional models DFN coupled to a two-dimensional problem for the boundary conditions. It was derived in \cite{sulzer_part_I} from a fully three-dimensional DFN model in the limit of large current collector conductivity and small aspect ratio.
A further limit, in which the current collector conductivity is even larger, lead to a model in which only a single through-cell one-dimensional problem needs to be solved, with an additional two-dimensional problem needed to calculate an in-series resistance. We refer to this simpler model as the ``DFNCC'', where the ``CC'' stands for current collectors.

The reduced models in \cite{sulzer_part_I} were derived using the DFN model to describe the electrochemistry. However the simplifications to the transverse behaviour are independent of the choice of model for the through-cell behaviour. In the following we summarise the simplifications discussed in \cite{sulzer_part_I}.
We then show that, in the limit in which the current collector conductance is extremely large, the effect of the current collectors can be neglected entirely, and we recover the standard one-dimensional description of cell behaviour at the macroscale.

\subsection{The 2+1D Model}
In the limit in which the aspect ratio is small, the effective current collector conductance is large, and the cell cooling is moderate, the full 3D pouch cell model reduces to a collection of through-cell models coupled via two-dimensional problems in the transverse direction for the distribution of potential in the current collectors and the balance of energy. In terms of dimensional parameters, this model is suitable in the regime 
  \begin{equation*}
    \frac{L_x}{L_\perp} \ll 1, \quad \frac{I\ts{app}F}{L\ts{cn}\sigma\ts{ck}RT_{\infty}} \sim 1, \quad \frac{h\ts{face}L_\perp^2}{ \lambda\ts{eff}L_x^3} \sim 1,
    \quad \frac{h\ts{edge}L_\perp}{\lambda\ts{eff}L_x ^2} \sim 1,
  \end{equation*}
where $L_\perp$ is a typical transverse dimension (e.g. $L_y$, $L_z$ or $(L_yL_z)^{1/2})$, $\sigma\ts{ck}$ is the current collector conductivity ($\kin{n,p})$, $R$ is the molar gas constant, $T_\infty$ the ambient temperature, $I\ts{app}$ is the applied current, and $\lambda\ts{eff}$ the effective thermal conductivity. The heat transfer coefficient $h$ may in general be a function of space (e.g. for tab cooling \cite{hunt2016}), and we allow for the situation where the coefficient on the edges $h\ts{edge}$ (which acts on an area of typical size $L_x L_\perp$) may be larger than that on the faces $h\ts{face}$ (which acts on an area of typical size $L_\perp^2$).

In this limit the model is the two-dimensional pair-potential problem
\begin{subequations}
  \label{eqn:high_summary}
  
\begin{equation}
    \di{L\ts{cn}}\di{\sigma\ts{cn}}\di{\nabla_\perp}^2 \di{\phi\ts{s,cn}} = \di{\mathcal{I}}, 
    \qquad
    \di{L\ts{cp}}\di{\sigma\ts{cp}}\di{\nabla_\perp}^2 \di{\phi\ts{s,cp}} = -\di{\mathcal{I}},
    \qquad \mbox{ in }  \di{\Omega}
\end{equation}
with boundary conditions
\begin{equation}
    \di{\phi\ts{s,cn}} = 0 
    \quad \mbox{ on } 
    \partial \di{\Omega\ts{tab,cn,$\perp$}}, 
    \qquad
    \di{\nabla_\perp}\di{\phi\ts{s,cn}}\cdot\boldsymbol{n} = 0 \quad \mbox{ on } \partial \di{\Omega}\ts{ext,cn,$\perp$}
\end{equation}
\begin{equation}
    -\di{\sigma\ts{cp}} \di{\nabla_\perp}\di{\phi\ts{s,cp}}\cdot\boldsymbol{n} = \frac{\di{I_{\text{app}}}}{\di{A\ts{tab,cp}}}  \  \mbox{ on } \partial\di{\Omega}\ts{tab,cp,$\perp$},\qquad
    \di{\nabla_\perp}\di{\phi\ts{s,cp}}\cdot\boldsymbol{n} = 0 \  \mbox{ on } \partial\di{\Omega}\ts{ext,cp,$\perp$},
\end{equation}
where $\mathcal{I}(\phi\ts{s,cn}, \phi\ts{s,cp}, T)$ is the through-cell current density, given by any one-dimensional electrochemical model. This is coupled to the two-dimensional thermal problem
\begin{align}
 \di{\rho\ts{eff}} \pdv{\di{T}}{\di{t}} &=  \di{\lambda}\ts{eff}\di{\nabla_{\perp}}^2\di{T} + \di{\bar{Q}} - \frac{ (\di{h\ts{cn}} + \di{h\ts{cp}})}{\di{L}} (\di{T} - \di{T_{\infty}}), \qquad \mbox{ in } \Omega\\
 - \lambda\ts{eff} \nabla_\perp T \cdot \boldsymbol{n} &=  h\ts{eff} (T-T_{\infty}) \quad \mbox{ on }\partial\Omega
\end{align}
with initial condition $T = T_0$, where the x-averaged heat source is
\begin{align}
  \di{\bar{Q}} & = \bar{\mathcal{Q}}  + \frac{\di{L\ts{cn}}}{\di{L}}  \di{\sigma\ts{cn}}\di{|\nabla_\perp}\di{\phi\ts{s,cn}}|^2 + \frac{\di{L\ts{cp}}}{\di{L}}  \di{\sigma\ts{cp}}\di{|\nabla_\perp}\di{\phi\ts{s,cp}}|^2,
\end{align}
with $\di{\bar{\mathcal{Q}}}$ being the x-averaged heat source from the one-dimensional electrochemical model with the current density given by $\mathcal{I} = I\ts{app}/(L_yL_z)$. In the above $I\ts{app}$ is the applied current, $A\ts{tab,cp}$ is the positive tab area, $\rho\ts{eff}$ is the effective volumetric heat capacity, $L$ is the total cell thickness ($L_x + L\ts{cn} + L\ts{cp}$), $h\ts{eff}$ the effective edge heat transfer coefficient, and $h\ts{cn}$ and $h\ts{cp}$ are the heat transfer coefficients on the faces of the negative and positive current collectors, respectively. The effective edge heat transfer coefficient is
\[
h\ts{eff} = \frac{1}{(L_x+L\ts{cn}+L\ts{cp})} \int_{-L\ts{cn}}^{L_x+L\ts{cp}} h\ts{edge}   \, \dd{x},
\]
and the effective volumetric heat capacity  and thermal conductivity are
\begin{equation}
    \rho\ts{eff} = \frac{\sum\limits\ts{k} \rho\ts{k} c\ts{p,k} L\ts{k}}{\sum\limits\ts{k} L\ts{k}}, \qquad \lambda\ts{eff} = \frac{\sum\limits\ts{k} \lambda\ts{k} L\ts{k}}{\sum\limits\ts{k} L\ts{k}},
\end{equation}
respectively. Here $\rho\ts{k}$, $c\ts{p,k}$ and $\lambda\ts{k}$ are the the density, specific heat capacity and thermal conductivity, respectively, of each component.

\end{subequations}

\subsection{The very large conductance limit: CC Model}
In the limit where the current collector conductance is even larger, the potential in the current collectors is approximately uniform, and it suffices to solve a single volume-averaged through-cell model, rather than a collection of through-cell models. Further, if the edge heat transfer coefficient $h\ts{eff}$ is suitably small, and the face heat transfer coefficients $h\ts{cn}$ and $h\ts{cp}$ are uniform in space, the cell temperature is approximately uniform. In terms of dimensional parameters, this simplified model is suitable when
  \begin{equation*}
    \frac{L_x}{L_\perp} \ll 1, \quad \frac{I\ts{app}F}{L\ts{cn}\sigma\ts{ck}RT_{\infty}} \ll 1, \quad \frac{h\ts{face}L_\perp^2}{ \lambda\ts{eff}L_x^3} \sim 1,
    \quad \frac{h\ts{edge}L_\perp}{\lambda\ts{eff}L_x ^2} \ll 1,
  \end{equation*}
The ``CC'' model consists of an algebraic expression for the terminal voltage and a single differential equation for the average cell temperature 
\begin{subequations}
    \label{model_very_high}
    \begin{align}
  \label{eq:very_high_V}
  \di{V} & =  \di{\mathcal{V}}\left(\di{\mathcal{I}}, \YZav{\di{T}}\right)-\di{R\ts{cp}} \di{I\ts{app}}- \di{R\ts{cn}} \di{I\ts{app}},\\
  \label{eq:very_high_T}
\di{\rho\ts{eff}} \pdv{\YZav{\di{T}}}{\di{t}} &=    \di{\bar{\mathcal{Q}}}\left(\di{\mathcal{I}}, \YZav{\di{T}}\right) - \frac{(\di{h}\ts{cn}+\di{h}\ts{cp})}{\di{L}} (\YZav{\di{T}}-\di{T}_{\infty})- \frac{(\YZav{\di{T}}-T_\infty)}{\di{L_y} \di{L_z}} \int_{\partial \di{\Omega}}\di{{h\ts{eff}}}\, \dd{\di{s}} \\ \nonumber
  & \hspace{3cm} \mbox{ }+  \di{H}\ts{cn} \di{I\ts{app}}^2+ \di{H\ts{cp}} \di{I\ts{app}}^2 \qquad \mbox{ in }\di{\Omega},
\end{align}
where $\mathcal{V}$ and $\bar{\mathcal{Q}}$ are the voltage and $x$-averaged heat source terms determined from the solution of a single one-dimensional electrochemical model. The current collector resistances are computed as 
\begin{equation} 
\di{R\ts{cn}} = \frac{ \YZav{\di{f\ts{n}}}}{\di{L_y} \di{L_z} \di{L\ts{cn}}\di{\sigma\ts{cn}}}, 
\qquad 
\di{R\ts{cp}} =  \frac{1}{\di{L_y} \di{L_z} \di{\sigma\ts{cp}}\di{A\ts{tab,cp}}}\int_{\partial\di{\Omega}\ts{tab,cp,$\perp$}}   \di{f}\ts{p}\, \dd{\di{s}}
\end{equation}
and the coefficients related to Ohmic heating in the current collectors are
\begin{equation}
\di{H}\ts{cn} = \frac{\di{L}\ts{cn}\YZav{\vert \nabla_\perp \di{f}\ts{n} \vert^2}}{\di{L}(\di{L_y} \di{L_z} \di{L\ts{cn}})^2 \di{\sigma}\ts{cn}} , \qquad \di{H}\ts{cp} = \frac{\di{L}\ts{cp}\YZav{\vert \nabla_\perp \di{f}\ts{p} \vert ^2}}{\di{L}(\di{L_y} \di{L_z} \di{L\ts{cp}})^2 \di{\sigma}\ts{cp}}  ,
\end{equation}
where $f\ts{n}$ and $f\ts{p}$ satisfy the auxilliary equations
\begin{equation}
    \di{\nabla_\perp}^2 \di{f\ts{n}}=  -1, \qquad  \qquad 
   \di{\nabla_\perp^2} \di{f\ts{p}} = 1 \qquad \mbox{ in }  \Omega,
\end{equation}
\begin{align}
  \di{f}\ts{n} &= 0 \quad \mbox{ on } \partial \di{\Omega}\ts{tab,cn,$\perp$},& 
\di{\nabla_\perp} \di{f\ts{n}}\cdot\boldsymbol{n} &= 0 \quad \mbox{ on } \partial \di{\Omega}\ts{ext,cn,$\perp$}  ,\\
\di{\nabla_\perp} \di{f\ts{p}}\cdot\boldsymbol{n} &= \frac{\di{L_y} \di{L_z}\di{L\ts{cp}}}{\di{A\ts{tab,cp}}}  \  \mbox{ on } \partial\di{\Omega}\ts{tab,cp,$\perp$},&
\di{\nabla_\perp} \di{f\ts{p}}\cdot\boldsymbol{n} &= 0 \
\mbox{ on } \partial\di{\Omega}\ts{ext,cp,$\perp$}, \qquad \YZav{\di{f\ts{p}}}=0.
\end{align}
The potential distribution in the current collectors can be determined from $f\ts{n}$ and $f\ts{p}$ via
\begin{equation}
    \label{eq:very_high_phi_cc}
    \di{\phi\ts{s,cn}} = -\frac{\di{I\ts{app}} \di{f}\ts{n}}{\di{L_y}\di{L_z}\di{L\ts{cn}}\di{\sigma\ts{cn}}}, 
    \qquad
    \di{\phi\ts{s,cp}} = \di{V} + \frac{\di{I\ts{app}}\di{f\ts{p}}}{\di{L_y}\di{L_z}\di{L\ts{cp}}\di{\sigma\ts{cp}}}.
\end{equation}
\end{subequations}

An approach sometimes used in the literature is to retain the spatial derivatives in the energy balance equation, but use heat source terms from the averaged one-dimensional electrochemical model (see e.g.~\cite{hosseinzadeh2018}). Such an approach corresponds to replacing
$\bar{\mathcal{Q}}(\mathcal{I}, T)$ with $\bar{\mathcal{Q}}(\mathcal{I}, \YZav{T})$
and 
replaces \eqref{eq:very_high_T} with 
\begin{subequations}
\label{eqn:very_high_T_spatial}
\begin{align}
 \di{\rho\ts{eff}} \pdv{\di{T}}{\di{t}} &=  \di{\lambda}\ts{eff}\di{\nabla_{\perp}}^2\di{T} + \di{\bar{Q}}(\mathcal{I}, \YZav{T}) - \frac{ (\di{h\ts{cn}} + \di{h\ts{cp}})}{\di{L}} (\di{T} - \di{T_{\infty}}) \\
 \nonumber 
 & \hspace{1cm} \mbox{ } +\frac{\di{L\ts{cn}}}{\di{L}}  \di{\sigma\ts{cn}}\di{|\nabla_\perp}\di{\phi\ts{s,cn}}|^2 + \frac{\di{L\ts{cp}}}{\di{L}}  \di{\sigma\ts{cp}}\di{|\nabla_\perp}\di{\phi\ts{s,cp}}|^2, \qquad \mbox{ in }\Omega \\
 -\lambda\ts{eff}\nabla_\perp T \cdot \boldsymbol{n} &=  h\ts{eff}( T-T_\infty) \quad \mbox{ on }\partial\Omega
\end{align}
\end{subequations}
This model captures the variation due to Ohmic heating in the current collectors and  cooling at the boundaries, but neglects the spatial variation of the  heat source within the cell. In the following results we use \eqref{eqn:very_high_T_spatial} in favour of \eqref{eq:very_high_T}.

\subsection{The extremely large conductance limit: 0D Model}
In the CC model  the conductivity of the current collectors is high enough that the potential is approximately uniform across them, and the resistances due to the current collectors are calculated as a perturbation. If this perturbation is so small as to be negligible, these correction terms can be ignored. In that case the potential in the current collectors is uniform 
\begin{equation}
    \phi\ts{s,cn} = 0, \quad \phi\ts{s,cp} = V,
\end{equation}
and we arrive at a model in which the effects of the current collectors are ignored entirely. In this model the cell behaviour is uniform in $(y,z)$, and the temperature, which is now a function of time only, is governed by the ODE
    \begin{align}
  \label{eq:extremelt_high_T}
\di{\rho\ts{eff}} \pdv{\di{T}}{\di{t}} &=    \di{\bar{\mathcal{Q}}}\left(\di{\mathcal{I}}, \di{T}\right) - \frac{(\di{h}\ts{cn}+\di{h}\ts{cp})}{\di{L}} (\di{T}-\di{T}_{\infty})- \frac{(\di{T}-T_\infty)}{\di{L_y} \di{L_z}} \int_{\partial \di{\Omega}}\di{h\ts{eff}}\, \dd{\di{s}} \mbox{ in }\di{\Omega},
\end{align}
    with initial condition $T=T_0$. The CC model includes terms of order $({I\ts{app}F}/{L\ts{cn}\sigma\ts{ck}RT_{\infty}})$ but neglects those of order  $({I\ts{app}F}/{L\ts{cn}\sigma\ts{ck}RT_{\infty}})^2$ , while   the 
    $0$D model neglects terms of order   
    $({I\ts{app}F}/{L\ts{cn}\sigma\ts{ck}RT_{\infty}})$.
   
\section{Through-cell models}\label{sec:through-cell-models}
Here we give the governing equations for the DFN, SPMe and SPM. In \cite{Marquis2019} it was shown that the SPM and SPMe can be derived from the DFN via asymptotic analysis in the limit in which the electrical conductivity in the electrodes and electrolyte is large and the timescale
for the migration of lithium ions in the electrolyte is small relative to the typical timescale of a
discharge. For clarity, we outline the requirements for the simplified models to be valid, but we omit the details of the derivation. 

\subsection{DFN}
The one-dimensional DFN model comprises equations for charge conservation
\begin{subequations}\label{eqn:DFN}
\begin{align}
    \label{eqn:DFN:solid:ohms}
    \di{\mathcal{I}}-\di{i\ts{e,k}} &= - \di{\sigma\ts{k}} \pdv{\di{\phi\ts{k}}}{\di{x}}, &&\kin{n, p} \\
    \label{eqn:DFN:electrolyte:current1}
    \pdv{\di{i\ts{e,k}}}{\di{x}} &= \begin{cases}
                        \di{a\ts{k}}\di{j\ts{k}}, \quad \text{k} = \text{n, p}\\
                        0, \qquad \text{k} = \text{s}
                        \end{cases},
                        && \\
    \label{eqn:DFN:electrolyte:current2}
    \di{i\ts{e,k}} &= \epsilon\ts{k}^{\text{b}} \di{\kappa\ts{e}}(\di{c\ts{e,k}}, \di{T}) \left( -  \pdv{\di{\phi\ts{e,k}}}{\di{x}} + 2(1-t^+)\frac{\di{R}\di{T}}{\di{F}} \pdv{\di{x}}( \log(\di{c\ts{e,k}}))\right), && \kin{n, s, p}; \\
    \intertext{mass conservation}
    \label{eqn:DFN:c_s_k}
       \pdv{\di{c\ts{s,k}}}{\di{t}} &= -\frac{1}{\di{r}^2}\pdv{\di{r}}\bigg(\di{r}^2 \di{N\ts{s,k}}\bigg), \quad \di{N\ts{s,k}} = -\di{D\ts{s,k}}(\di{c\ts{s,k}}, \di{T\ts{k}})\pdv{\di{c\ts{s,k}}}{\di{r}}, \quad && \kin{n, p}, \\
    \label{eqn:DFN:c_e_k}
     \epsilon\ts{k} \pdv{\di{c\ts{e,k}}}{\di{t}} &= -\pdv{\di{N\ts{e,k}}}{\di{x}} + \frac{1}{F} \pdv{\di{i\ts{e,k}}}{\di{x}}, \quad && \kin{n, s, p},\\
    \label{eqn:DFN:N_e_k}
      \di{N\ts{e,k}} &= -\epsilon^{\text{b}}\ts{k} \di{D\ts{e}}(\di{c\ts{e,k}}, \di{T\ts{k}}) \pdv{\di{c\ts{e,k}}}{\di{x}} + \frac{t^+}{\di{F}} \di{i\ts{e,k}}, \quad && \kin{n, s, p} \\
    \intertext{and electrochemical reactions}
      \label{eqn:N+1D-DFN:j_k}
     \di{j\ts{k}} &= \di{j\ts{$0,$k}} \sinh\left(\frac{\di{F}\di{\eta\ts{k}}}{2\di{R_g}\di{T}\ts{k}} \right),
     &&\kin{n, p}, \\
     \label{eqn:DFN:j0_k}
    \di{j\ts{$0,$k}} &= \di{m\ts{k}}(\di{T\ts{k}}) (\di{c\ts{s,k}})^{1/2} (\di{c\ts{s,k,max}} - \di{c\ts{s,k}})^{1/2}(\di{c\ts{e,k}})^{1/2} \big|_{\di{r}=\di{R}\ts{k}}  && \kin{n, p},\\
      \label{eqn:DFN:eta_k}
     \di{\eta\ts{k}} &= \di{\phi\ts{s,k}} - \di{\phi\ts{e,k}} - \di{U\ts{k}}(\di{c\ts{s,k}}, \di{T})\big|_{\di{r\ts{k}}=\di{R\ts{k}}},  && \kin{n, p};
\end{align}
along with boundary conditions relating to charge conservation
\begin{align}
&\di{\phi}\ts{s,n}\big|_{\di{x}=0} = \di{\phi}\ts{s,cn}, \quad \di{\phi}\ts{s,p}\big|_{\di{x}=\di{L_x}}=\di{\phi}\ts{s,cp}, \\
\label{bc:DFN:no_electrolyte_cc_current}
&\di{i}\ts{e,n}\big|_{\di{x}=0} = 0, \quad \di{i}\ts{e,p}\big|_{\di{x}=\di{L_x}}=0, \\
\label{bc:DFN:charge_continuity_Ln}
&\di{\phi}\ts{e,n}\big|_{\di{x}=\di{L\ts{n}}} = \di{\phi}\ts{e,s}\big|_{\di{x}=\di{L\ts{n}}}, \quad \di{i}\ts{e,n}\big|_{\di{x}=\di{L\ts{n}}} = \di{i}\ts{e,s}\big\vert_{\di{x}=\di{L\ts{n}}}, \\
\label{bc:DFN:charge_continuity_Lp}
&\phi\ts{e,s}\big|_{\di{x}=\di{L_x}-\di{L\ts{p}}} = \di{\phi}\ts{e,p}\big|_{\di{x}=\di{L_x}-\di{L\ts{p}}}, \quad
\di{i}\ts{e,s}\big|_{\di{x}=\di{L_x}-\di{L\ts{p}}} = \di{i}\ts{e,p}\big|_{\di{x}=\di{L_x}-\di{L}\ts{p}};
\intertext{mass conservation in the electrolyte}
\label{bc:DFN:no_electrolyte_cc_flux}
&\di{N}\ts{e,n}\big|_{\di{x}=0} = 0, \quad \di{N}\ts{e,p}\big|_{\di{x}=\di{L_x}}=0,  \\
\label{bc:DFN:concentration_continuity_Ln}
&\di{c}\ts{e,n}\big|_{\di{x}=\di{L}\ts{n}} = \di{c}\ts{e,s}|_{\di{x}=\di{L}\ts{n}}, \quad \di{N}\ts{e,n}\big|_{\di{x}=\di{L}\ts{n}}=\di{N}\ts{e,s}\big|_{\di{x}=\di{L}\ts{n}}, \\
\label{bc:DFN:concentration_continuity_Lp}
&\di{c}\ts{e,s}|_{\di{x}=\di{L_x}-\di{L}\ts{p}}=\di{c}\ts{e,p}|_{\di{x}=\di{L_x}-\di{L}\ts{p}}, \quad \di{N}\ts{e,s}\big|_{\di{x}=\di{L_x}-\di{L}\ts{p}}=\di{N}\ts{e,p}\big|_{\di{x}=\di{L_x}-\di{L}\ts{p}};
\intertext{and mass conservation in the electrode active material}
\label{bc:DFN:particle}
&\di{N}\ts{s,k}\big|_{\di{r}\ts{k}=0} = 0, \quad \di{N}\ts{s,k}\big|_{\di{r}\ts{k}=\di{R\ts{k}}} = \frac{\di{j}\ts{k}}{\di{F}}, \quad \kin{n, p}.
\end{align}
\end{subequations}
In addition, the following initial conditions are prescribed for the lithium concentrations in the solid and electrolyte
\begin{subequations}
    \begin{align}
    \label{ic:DFN:solid}
	\di{c}\ts{s,k}(\di{x},\di{y},\di{z},\di{r},0) &= \di{c}\ts{s,k,0}, && \kin{n, p},\\
	\label{ic:DFN:electrolyte}
	\di{c}\ts{e,k}(\di{x},\di{y},\di{z},0) &= \di{c}\ts{e,0}, && \kin{n, s, p}.
    \end{align}
\end{subequations}

The heat source term $\mathcal{Q}\ts{k}$ is computed as 
\begin{equation}
\label{eq:Q_DFN}
 \mathcal{Q}\ts{k} = Q\ts{Ohm,k} + Q\ts{rxn,k} + Q\ts{rev,k},
\end{equation}
and accounts for Ohmic heating $Q\ts{Ohm,k}$ due to resistance in the solid and electrolyte, irreverisble heating due to electrochemical reactions $Q\ts{rxn,k}$, and reversible heating due to entropic changes in the the electrode $Q\ts{rev,k}$ \cite{bernardi1985}. In the electrodes these terms are computed as
\begin{align}
    \label{eq:Q_ohm}
    Q\ts{Ohm,k} &= - \left(i\ts{s,k} \pdv{\phi\ts{s,k}}{x} + i\ts{e,k} \pdv{ \phi\ts{e,k}}{x}\right), && \kin{n, p},\\
    \label{eq:Q_rxn}
    Q\ts{rxn,k} &= a\ts{k} j\ts{k} \eta\ts{k}, && \kin{n, p}, \\
    \label{eq:Q_rev}
    Q\ts{rev,k} &= a\ts{k} j\ts{k} T \pdv{U\ts{k}}{T}\bigg|_{T=T_\infty}, && \kin{n, p}.
\end{align}
However, in the separator there is no heat generation due to electrochemical effects, and we need only consider the Ohmic heat generation term given by
\begin{equation}
\label{eq:Q_sep}
 Q\ts{Ohm,s} = -\boldsymbol{i}\ts{e,s}\cdot\pdv{\phi\ts{e,s}}{x}.
\end{equation}
The x-averaged through-cell heat generation is given by
\begin{equation}\label{eq:Q_bar}
    \bar{\mathcal{Q}} = \frac{1}{L}\sum_{k} L\ts{k}\mathcal{Q}\ts{k}.
\end{equation}

\subsection{SPMe}
The one-dimensional single particle model with electrolyte (SPMe) was derived in \cite{Marquis2019} by employing asymptotic methods. The  limit taken is that of a short timescale for lithium-ion diffusion in the electrolyte relative to the discharge timescale, alongside high electron conductivity in the electrodes and high ion conductivity in the electrolyte. The conditions for application are summarised in Table~\ref{tab:conditions}. The model includes terms of
order $\mathcal{C}\ts{e}$, but neglects those of order $\mathcal{C}\ts{e}^2$, where $\mathcal{C}\ts{e}$ is defined in Table~\ref{tab:conditions}.
\begin{table}[h]
	\centering 
	\begin{tabular}{c l p{5cm}} 
  \toprule
  Parameter combination & Required size & Interpretation\\
  \midrule
  $\mathcal{C}\ts{e} = {\cal I}\ts{typ} L/(D\ts{e,typ}Fc\ts{n,max})$  & $\ll 1$ &Lithium-ion migration timescale is small relative to typical discharge timescale \\
  $R T \sigma\ts{k} / (F {\cal I}\ts{typ} L)$  & $\gg 1$ & Thermal voltage is large relative to the typical potential drop in electrode k\\
  $R T \kappa\ts{e,typ}/(F {\cal I}\ts{typ} L)$ & $\gg 1$ & Thermal voltage is large relative to the typical potential drop in the electrolyte \\
  $(R\ts{k})^2 {\cal I}\ts{typ}/(D\ts{s,k} F c\ts{n,max} L)$ & $\ll 1/\mathcal{C}\ts{e}$ & Solid diffusion occurs on a shorter or similar timescale to a discharge  \\
  ${\cal I}\ts{typ}/(m\ts{k} a\ts{k} (c\ts{e,typ})^{1/2} c\ts{n,max} L)$ 
  & $\ll 1/\mathcal{C}\ts{e}$ & Reactions occur on a shorter or similar timescale to a discharge \\
  \bottomrule
  \end{tabular}
  \caption{The key conditions to be satisfied for the application of the SPMe, with ${\cal I}\ts{typ} =  I\ts{app}/L_y / L_z$. In addition, it is required that $\mathcal{C}\ts{e} \ll L\ts{k}/L \ll 1/\mathcal{C}\ts{e}$, $\mathcal{C}\ts{e}\ll c\ts{p,max}/c\ts{n,max}\ll 1/\mathcal{C}\ts{e}$, and $\mathcal{C}\ts{e} \ll c\ts{e,typ}/c\ts{n,max}\ll 1/\mathcal{C}\ts{e}$, which are true in practical situations.}
  \label{tab:conditions}
\end{table}

In this paper we use a slightly modified version of the SPMe from that introduced in \cite{Marquis2019}. These modifications are discussed in Appendix~\ref{app:spme-mods}. The SPMe comprises equations for mass conservation in a single  ``average'' particle in each electrode
  \begin{gather}
       \pdv{\di{c\ts{s,k}}}{\di{t}} = -\frac{1}{\di{r}^2}\pdv{\di{r}}\bigg(\di{r}^2 \di{N\ts{s,k}}\bigg), \quad \di{N\ts{s,k}} = -\di{D\ts{s,k}}(\di{c\ts{s,k}}, \di{{T}\ts{k}})\pdv{\di{{c}\ts{s,k}}}{\di{r}}, \quad  \kin{n, p}, \\
        \di{{N}}\ts{s,k}\big|_{\di{r}\ts{k}=0} = 0, \quad \kin{n, p}, \quad \di{{N}}\ts{s,k}\big|_{\di{r}\ts{k}=\di{R\ts{k}}} = 
        \begin{cases}
	 \displaystyle	  \frac{\di{\mathcal{I}}}{\di{F} \di{a\ts{n}} \di{L\ts{n}}}, \quad &\text{k}=\text{n}, \\[3mm] 
	 \displaystyle	  -\frac{\di{\mathcal{I}}}{\di{F} \di{a\ts{p}} \di{L\ts{p}}}, \quad &\text{k}=\text{p},
        \end{cases} , \\
    	\di{{c}}\ts{s,k}(\di{y},\di{z},\di{r},0) = \di{c}\ts{s,k,0}, \quad \kin{n, p};
  \end{gather}
  and mass conservation in the electrolyte
\begin{gather}
     \epsilon\ts{k}\pdv{\di{c\ts{e,k}}}{\di{t}} = -\pdv{\di{N\ts{e,k}}}{\di{x}} + 
    \begin{cases} \displaystyle
    \frac{\di{\mathcal{I}}}{\di{F}\di{L\ts{n}}}, \quad &\text{k}=\text{n}, \\ 
    0, \quad &\text{k}=\text{s}, \\ 
 \displaystyle   -\frac{\di{\mathcal{I}}}{\di{F}\di{L\ts{p}} }, \quad &\text{k}=\text{p},
    \end{cases}
   , \\ 
    \label{eqn:SPMe:N_e_k}
     \di{N\ts{e,k}} = -\epsilon\ts{k}^{\text{b}} \di{D\ts{e}}(\di{c\ts{e,k}}, \di{T}) \pdv{\di{c\ts{e,k}}}{\di{x}} +     
	\begin{cases} 
 \displaystyle	  \frac{\di{x} t^+ \di{\mathcal{I}}}{\di{F}\di{L\ts{n}}}, \quad &\text{k}=\text{n}, \\[3mm] 
   \displaystyle   \frac{t^+\di{\mathcal{I}}}{\di{F}}, \quad &\text{k}=\text{s}, \\[3mm] 
 \displaystyle	 \frac{(\di{L}-\di{x})t^+ \di{\mathcal{I}}}{\di{F} \di{L\ts{p}}}, \quad &\text{k}=\text{p},
    \end{cases}\\
     \di{N\ts{e,n}}\big|_{\di{x}=0} = 0, \quad \di{N\ts{e,p}}\big|_{\di{x}=\di{L}}=0,  \\ 
    \di{c\ts{e,n}}|_{\di{x}=\di{L\ts{n}}}=\di{c\ts{e,s}}|_{\di{x}=\di{L\ts{n}}}, \quad \di{N\ts{e,n}}\big|_{\di{x}=\di{L\ts{n}}}=\di{N\ts{e,s}}\big|_{\di{x}=\di{L\ts{n}}}, \\
    \di{c\ts{e,s}}|_{\di{x}=\di{L}-\di{L\ts{p}}}=\di{c\ts{e,p}}|_{\di{x}=\di{L}-\di{L\ts{p}}}, \quad \di{N\ts{e,s}}\big|_{\di{x}=\di{L}-\di{L\ts{p}}}=\di{N\ts{e,p}}\big|_{\di{x}=\di{L}-\di{L\ts{p}}} \\
	\di{c}\ts{e,k}(\di{x},\di{y},\di{z},0) = \di{c}\ts{e,0}, \quad \kin{n, s, p}.
\end{gather}

The through-cell current density, $\di{\mathcal{I}}$, and local voltage $\di\mathcal{V} = \phi\ts{s,cp} - \phi\ts{s,cn}$ are related through
\begin{equation}\label{eqn:SPMe:voltage}
 \di{\mathcal{V}} = \di{{U}_{\text{eq}}} +\di{{\eta}_r}
     +  \di{{\eta}_c} + \di{{\Delta\Phi}_{\text{Elec}}} + \di{{\Delta\Phi}_{\text{Solid}}},
\end{equation}
where ${U}\ts{eq}$ is the electrode-averaged open-circuit voltage
\begin{equation}\label{eqn:spm_Ueq}
      \di{{U}_{\text{eq}}} = \di{U\ts{p}}\left(\di{{c}\ts{s,p}}\big|_{\di{r}=\di{R\ts{p}}}, \di{T}\right) - \di{U\ts{n}}\left(\di{{c}\ts{s,n}}\big|_{\di{r}=\di{R\ts{n}}}, \di{T}\right), \\ 
\end{equation}
$\di{{\eta}\ts{r}}$ is the electrode-averaged reaction overpotential and is given by the difference of the average positive and negative electrode reaction overpotentials as 
\begin{gather}\label{eqn:spm_eta_r}
    \di{{\eta}\ts{r}} = \di{{\eta}\ts{p}} - \di{{\eta}\ts{n}} \\
    \di{{\eta}\ts{n}} = \frac{2\di{R}\di{T}}{\di{F}}\sinh^{-1}\left(\frac{\di{\mathcal{I}}}{\di{a\ts{n}}\di{{j}\ts{$0$,n}} \di{L\ts{n}}}\right), \\
    \di{{\eta}\ts{p}} = -\frac{2\di{R}\di{T}}{\di{F}}\sinh^{-1}\left(\frac{\di{\mathcal{I}}}{\di{a\ts{p}}\di{{j}\ts{$0$,p}} \di{L\ts{p}}}\right), 
\end{gather}
where $\di{j}\ts{$0$,n}$ and $\di{j}\ts{$0$,p}$ are the average negative and positive exchange current densities given by
\begin{gather}
	 \di{{j}\ts{$0$,n}} =  \frac{1}{\di{L\ts{n}}}\int_0^{\di{L\ts{n}}} \di{m\ts{n}}(\di{T}) (\di{c\ts{s,n}})^{1/2}(\di{c\ts{s,n,max}}-\di{c\ts{s,n}})^{1/2} (\di{c\ts{e,n}})^{1/2} \, \text{d}\di{x}, \\
	 \di{{j}\ts{$0$,p}} =  \frac{1}{\di{L\ts{p}}}\int_{\di{L}-\di{L\ts{p}}}^{\di{L}} \di{m\ts{p}}(\di{T})  (\di{c\ts{s,p}})^{1/2}(\di{c\ts{s,p,max}}-\di{c\ts{s,p}})^{1/2} (\di{c\ts{e,p}})^{1/2} \, \text{d}\di{x}.
\end{gather}
The remaining terms in \eqref{eqn:SPMe:voltage} are the electrolyte concentration overpotential
\begin{equation}\label{eqn:SPMe:eta_c}
	 \di{{\eta}_c} =  2(1-t^+)\frac{\di{R}\di{T}}{\di{F}}\log\left(\frac{\di{{c}\ts{e,p}}}{ \di{{c}\ts{e,n}}}\right), 
\end{equation}
the electrolyte Ohmic losses
\begin{equation}
   \di{{\Delta \Phi}_{\text{Elec}}}=
   -\frac{\di{\mathcal{I}}}{\di{\kappa\ts{e}}(\di{c\ts{e,0}}, \di{T})}\left(\frac{\di{L\ts{n}}}{3\epsilon\ts{n}^b} + \frac{\di{L\ts{s}}}{\epsilon\ts{s}^b} + \frac{\di{L\ts{p}}}{3\epsilon\ts{p}^b} \right),
\end{equation}
and the solid-phase Ohmic losses
\begin{equation}
	 \di{{\Delta \Phi}_{\text{Solid}}} =  -\frac{\di{\mathcal{I}}}{3}\left(\frac{\di{L\ts{p}}}{\di{\sigma\ts{p}}} + \frac{\di{L\ts{n}}}{\di{\sigma\ts{n}}} \right).
\end{equation}
If the through-cell current $\di{\mathcal{I}}$ is provided then we can \emph{a-posteriori} read $\mathcal{V}$ from (\ref{eqn:SPMe:voltage}). However, if $\mathcal{V}$ is provided then (\ref{eqn:SPMe:voltage}) provides an algebraic constraint that must be enforced during the solution process. 

The remaining key variables are given by the following algebraic expressions. The negative and positive electrode potentials are given by
\begin{gather}
    \di{\phi\ts{s,n}} = \di{\phi\ts{s,cn}} + \frac{\di{\mathcal{I}} \di{x}}{2 \di{\sigma\ts{n}} \di{L\ts{n}}}(2 \di{L\ts{n}} - \di{x}), \\
   \di{\phi\ts{s,p}} = \di{\phi\ts{s,cp}} + \frac{\di{\mathcal{I}} (\di{x}-\di{L_x})(\di{L_x} - 2 \di{L_p} - \di{x})}{2 \di{\sigma\ts{p}\di{L\ts{p}}}}; 
\end{gather}
the electrolyte potential is given by 
\begin{gather}
        \label{eqn:SPMe:phi_e_n}
		\di{\phi\ts{e,n}} = \di{\Phi\ts{e}} + 2(1-t^+)\frac{\di{R}\di{T}}{\di{F}}\log\left(\frac{\di{c\ts{e,n}}}{\di{{c}\ts{e,n}}} \right)
		- \frac{\di{\mathcal{I}}}{\di{\kappa\ts{e}}(\di{c\ts{e,0}},\di{T})}\left(\frac{\di{x}^2-\di{L\ts{n}}^2}{2 \epsilon\ts{n}^b \di{L\ts{n}}} + \frac{\di{L\ts{n}}}{\epsilon\ts{s}^b} \right), \\
        \label{eqn:SPMe:phi_e_s}
		\di{\phi\ts{e,s}} = \di{\Phi\ts{e}}  + 2(1-t^+)\frac{\di{R}\di{T}}{\di{F}}\log\left(\frac{\di{c\ts{e,s}}}{\di{{c}\ts{e,n}}}\right)
		- \frac{\di{\mathcal{I}} \di{x}}{ \di{\kappa\ts{e}}(\di{c\ts{e,0}}, \di{T})\epsilon\ts{s}^b},\\
        \label{eqn:SPMe:phi_e_p}
		\di{\phi\ts{e,n}} = \di{\Phi\ts{e}} + 2(1-t^+)\frac{\di{R}\di{T}}{\di{F}}\log\left(\frac{\di{c\ts{e,p}}}{\di{{c}\ts{e,n}}}\right)
		- \frac{\di{\mathcal{I}}}{ \di{\kappa\ts{e}}(\di{c\ts{e,0}}, \di{T})}\left(\frac{\di{x}(2 \di{L_x}-\di{x})+\di{L\ts{p}}^2-\di{L_x}^2}{2\epsilon\ts{p}^b \di{L\ts{p}}} + \frac{\di{L_x}-\di{L\ts{p}}}{\epsilon\ts{s}^b} \right),
\end{gather}
where
\begin{gather}
    \di{\Phi}\ts{e} = {\phi}\ts{s,n} - {\eta}\ts{n} + \frac{\di{\mathcal{I}} \di{L_n}}{\kappa\ts{e}(\di{c\ts{e,0}}, \di{T})}
    \left(
        \frac{1}{\epsilon\ts{n}^b} 
        -\frac{1}{\epsilon\ts{s}^b} 
    \right);
\end{gather}
and the average negative and positive interfacial current densities are given by
\begin{gather}
    \di{{j}}\ts{n} = \frac{\di{\mathcal{I}}}{\di{F} \di{a\ts{n}} \di{L\ts{n}}}, \qquad
    \di{{j}}\ts{p} = -\frac{\di{\mathcal{I}}}{\di{F} \di{a\ts{p}} \di{L\ts{p}}}.
\end{gather}
The heat source term $\mathcal{Q}$ is computed using \eqref{eq:Q_DFN}--\eqref{eq:Q_bar}.

\subsection{SPM}
The SPMe retains terms of order ${\cal C}\ts{e}$ but neglects terms of order ${\cal C}\ts{e}^2$. If ${\cal C}\ts{e}$ is so small that the terms of order ${\cal C}\ts{e}$ may themselves be neglected, the result is the one-dimensional single particle model (SPM) \cite{Marquis2019}. Thus, the SPM is less accurate than the SPMe, but also slightly simpler.
The SPM consists only of equations for mass conservation in an average particle in each electrode
\begin{gather}
       \pdv{\di{{c}\ts{s,k}}}{\di{t}} = -\frac{1}{\di{r}^2}\pdv{\di{r}}\bigg(\di{r}^2 \di{{N}\ts{s,k}}\bigg), \quad \di{{N}\ts{s,k}} = -\di{D\ts{s,k}}(\di{{c}\ts{s,k}}, \di{{T}\ts{k}})\pdv{\di{{c}\ts{s,k}}}{\di{r}}, \quad  \kin{n, p}, \\
        \di{{N}}\ts{s,k}\big|_{\di{r}\ts{k}=0} = 0, \quad  \kin{n, p},\quad  \di{{N}}\ts{s,k}\big|_{\di{r}\ts{k}=\di{R\ts{k}}} =  
        \begin{cases}\displaystyle
		  \frac{\di{\mathcal{I}}}{\di{F} \di{a\ts{n}} \di{L\ts{n}}}, \quad &\text{k}=\text{n}, \\[3mm]
	\displaystyle	  -\frac{\di{\mathcal{I}}}{\di{F} \di{a\ts{p}} \di{L\ts{p}}}, \quad &\text{k}=\text{p},
        \end{cases} \\
    	\di{{c}}\ts{s,k}(\di{y},\di{z},\di{r},0) = \di{c}\ts{s,k,0}, \quad \kin{n, p}.
\end{gather}

The through-cell current, $\di{\mathcal{I}}$, and local voltage $\di\mathcal{V}$ are related through
\begin{equation}\label{eqn:SPM:voltage}
 \di{\mathcal{V}} = \di{{U}_{\text{eq}}} +\di{{\eta}_r},
\end{equation}
where ${U}\ts{eq}$ and $\di{{\eta}\ts{r}}$ are given by \eqref{eqn:spm_Ueq} and \eqref{eqn:spm_eta_r}, respectively. 
The negative and positive electrode potentials are given by
\begin{equation}
    \di{\phi}\ts{s,n} = \di{\phi\ts{s,cn}}, \quad \di{\phi\ts{s,p}} = \di{\phi\ts{s,cp}},
\end{equation}
the electrolyte potential is given by
\begin{equation}
    \di{\phi\ts{e}} = \di{\phi\ts{s,n}} - \di{U\ts{n}}\left(\di{{c}\ts{s,n}}\big|_{\di{r}=\di{R\ts{n}}}, \di{T}\right)- \di{{\eta}\ts{n}},
\end{equation}
and the electrolyte concentration is given by
\begin{equation}
    \di{c}\ts{e,k} = \di{c\ts{e,0}}.
\end{equation}
Finally, the heat source term $\mathcal{Q}$ is computed using \eqref{eq:Q_DFN}--\eqref{eq:Q_bar}. However, since the potentials in the SPM are all constant in space, the only non-zero contributions to the heating come from the terms due to electrochemical reactions $Q\ts{rxn,k}$ and $Q\ts{rev,k}$. The Ohmic heating terms appear at higher order, and are included in the SPMe.

Although the SPM is simpler than the SPMe, we note that in terms of 
computational complexity they are similar---the SPMe essentially comprises three uncoupled one-dimensional partial differential equations in comparison to the two uncoupled one-dimensional partial differential equations of the SPM. In contrast the DFN model, as already noted, is pseudo-two-dimensional.

\section{Critical Comparison of Isothermal 1+1D Models}\label{sec:comparison}
In this section, we provide a numerical comparison of the models depicted in Figure~\ref{fig:model_schematic}. For ease of exposition we consider the case in which all variables are uniform in $y$, so that each model has  one-dimensional current collectors (in $z$) (in the ``extremely large conductance'' limit  the effects of the current collectors are ignored). In total we consider 9 models: 1+1D DFN, 1+1D SPM, 1+1D SPMe, DFNCC, SPMeCC, SPMCC, DFN, SPMe and SPM. Each model is implemented within the open source battery modelling software PyBaMM (Python Battery Mathematical Modelling)\cite{pybamm}. PyBaMM is a modular open-source  framework that allows users to build and solve physics-based models of lithium-ion cells by selecting  different physical effects for each component of the cell, so is well-suited to comparing a suite of different models like those in Figure~\ref{fig:model_schematic}. 
The model equations are discretised in space using the finite-volume method and integrated in time using an adaptive, variable-order backward differentiation
formula. For the spatial discretisation we use the following number of grid points: $N_z=30$ in the $z$-direction of the current collectors; $N_{r\ts{n}}=20$ and $N_{r\ts{p}}=20$ in $r$-direction of the negative and positive particles, respectively; and $N_{x\ts{n}}=35$, $N_{x\ts{s}}=20$, and $N_{x\ts{p}}=35$ in the $x$-direction negative electrode, separator and positive electrode, respectively. Unless otherwise stated, results are presented for a 1C constant current discharge. To demonstrate the relative computational complexity of each model we present the number of states in each model\footnote{Here we refer to the number of states required by part of the model that is integrated in time. That is, we ignore the states in the ``CC'' part of the problem which is solved ``offline'' and is separate from the time integration.} alongside the solve time in Table~\ref{tab:states-and-solve-times}. We observe a dramatic decrease in both memory requirements and solve times as the complexity of the model is decreased. For example, the SPMeCC requires around 0.2$\%$ of the memory required by the 1+1D DFN and is over 400 times faster to solve than the 1+1D DFN. 

\begin{table}[h]
    \centering
    \resizebox{\textwidth}{!}{
    \begin{tabular}{l  c c c c c c}
        Model & 1+1D DFN & 1+1D SPMe & 1+1D SPM & DFN(CC) & SPMe(CC) & SPM(CC) \\
        \midrule
        Number of states & 49561 & 3961 & 1261 & 1651 & 131 & 41 \\
        Solve time [ms] & 8308 & 1514 & 105 & 259 & 18 & 13 
    \end{tabular}
    }
    \caption{The number of states required for the time integration component of the isothermal version of each model and associated time integration solve time (i.e. excluding any initial time independent solves) for a \SI{1}{C} constant current disharge with $N_z=30$, $N_{x\ts{n}}=35$, $N_{x\ts{s}}=20$, $N_{x\ts{p}}=35$, $N_{r\ts{n}}=20$, and $N_{r\ts{p}}=20$. Since we have ignored the time-independent part of the problem, the results for the ``CC'' and 1D (no current collector effects) models are the same.}
    \label{tab:states-and-solve-times}
\end{table}

Despite the clear computational benefits of employing a reduced-order model, modelling errors are introduced and must be understood. To help quantify spatial and temporal errors, we use two measures. The first is the maximum absolute error at each point in $z$ defined for a variable $\Phi$ to be
\begin{equation}
    \epsilon_z = \max_{t} \left| \Phi\ts{1+1D DFN} - \Phi\ts{Reduced}  \right|,
\end{equation}
where $\Phi\ts{1+1D DFN}$ is the value of variable predicted by the 1+1D DFN model and $\Phi\ts{Reduced}$ is the value of the variable predicted by relevant the reduced-order model. The other measure of the error that we use is the maximum absolute error at each point in time in the discharge defined for each variable $\Phi$ as
\begin{equation}
   \epsilon_t = \max_{z} \left| \Phi\ts{1+1D DFN} - \Phi\ts{Reduced}  \right|.
\end{equation}
Both $\epsilon_z$ and $\epsilon_t$ automatically inherit the units of the variable, $\Phi$, under study. 

The results and conclusions presented below are specific to the parameter values in Tables~\ref{table:dimensional_parameter_values} and \ref{table:dimensional_coefficients}, and one should keep in mind that the most appropriate model is a function of the particular parameter values being used. In Section \ref{sec:comparison-summary}, we will discuss changes to key parameters that are likely to have a large impact upon the conclusions of this section.

\subsection{Comparison of terminal voltage}
\begin{figure}[htbp]
    \centering
    \includegraphics[width=0.9\textwidth]{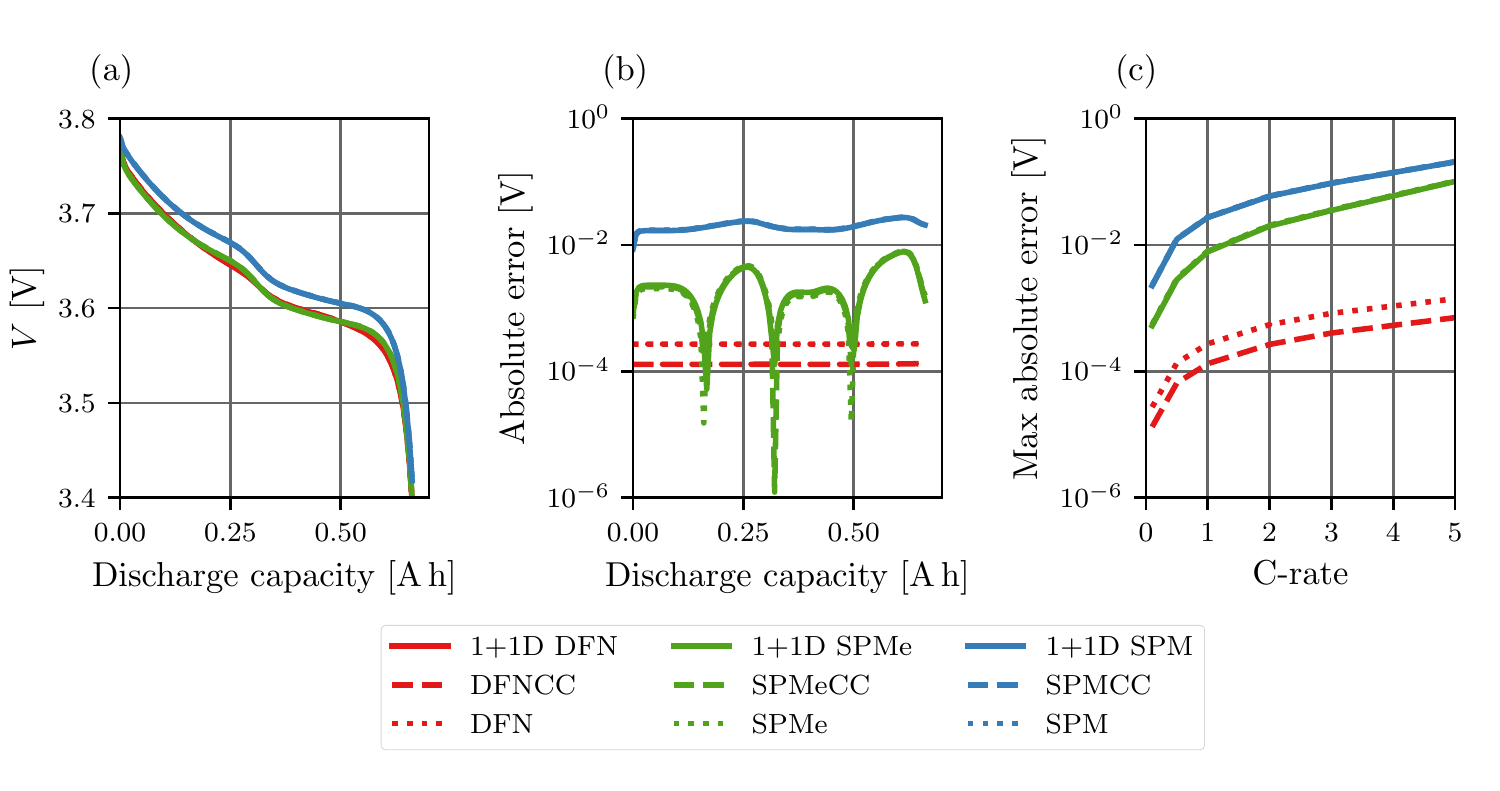}
    \caption{Comparison of the predicted terminal voltage: (a) $\SI{1}{C}$ discharge voltage profile predicted by each model; (b) $\SI{1}{C}$ discharge voltage profile absolute error between each reduced-order model and the 1+1D DFN at each time in the discharge; and (c) the maximum absolute error (across all times during the discharge) between each reduced-order model and the 1+1D DFN at a range of C-rates. The results for the 1+1D SPMe, SPMeCC, and SPMe are almost indistinguishable and the results for the 1+1D SPM, SPMCC, and SPM are indistinguishable.}\label{fig:terminal-voltage-comparison}
\end{figure}

The most commonly used output of lithium-ion battery models is typically the predicted terminal voltage. In Figure~\ref{fig:terminal-voltage-comparison} (a), and (b), we present the $\SI{1}{C}$ terminal voltage predicted by each model and the associated errors between the reduced-order model and the 1+1D DFN. At a glance, with the exception of the SPM, the reduced-order models are able to recover the behaviour predicted by the 1+1D DFN at $\SI{1}{C}$. However, in Figure~\ref{fig:terminal-voltage-comparison} (b), we observe between a one and two order of magnitude decrease in the error when using the DFN as the through-cell model, irrespective of the choice of transverse model. This suggests that simplifications in the the through-cell electrochemistry result in a greater increase in error than simplifications in transverse direction. In Figure~\ref{fig:terminal-voltage-comparison} (c), the maximum absolute error in the terminal voltage across the discharge is presented as a function of the C-rate. As expected, the error increases with increasing C-rate for all of the models considered. Additionally, the error in the models which use the SPM and SPMe increases more quickly with C-rate when compared with the models that use the DFN. This suggests that at higher C-rates retaining a more complex through-cell model becomes increasingly important. 

If the terminal voltage is the only quantity of interest then our results suggest that employing a reduced-order model that does not couple the current collector and electrochemical effects is the most appropriate model choice, in the sense the voltage can be accurately predicted at a greatly reduced computational cost. In particular, if one has a larger computational budget and has access to efficient differential algebraic equation (DAE) solvers then making use of the 1D DFN or DFNCC gives the best prediction of the voltage. However, if the  computational budget is more limited the 1D SPMe or SPMeCC may be more appropriate. Finally, we note that the errors between experimental data and more complex models (i.e. DFN) can be on the order of $10^{-1}\SI{}{V}$ \cite{ecker2015}, so in practice introducing a modelling error on the order of $10^{-2}\SI{}{V}$, such as in the SPMeCC, may not affect the ability of the model to replicate experimental results. 


\subsection{Comparison of particle concentrations}
\begin{figure}[htbp]
    \centering
    \includegraphics[width=0.8\textwidth]{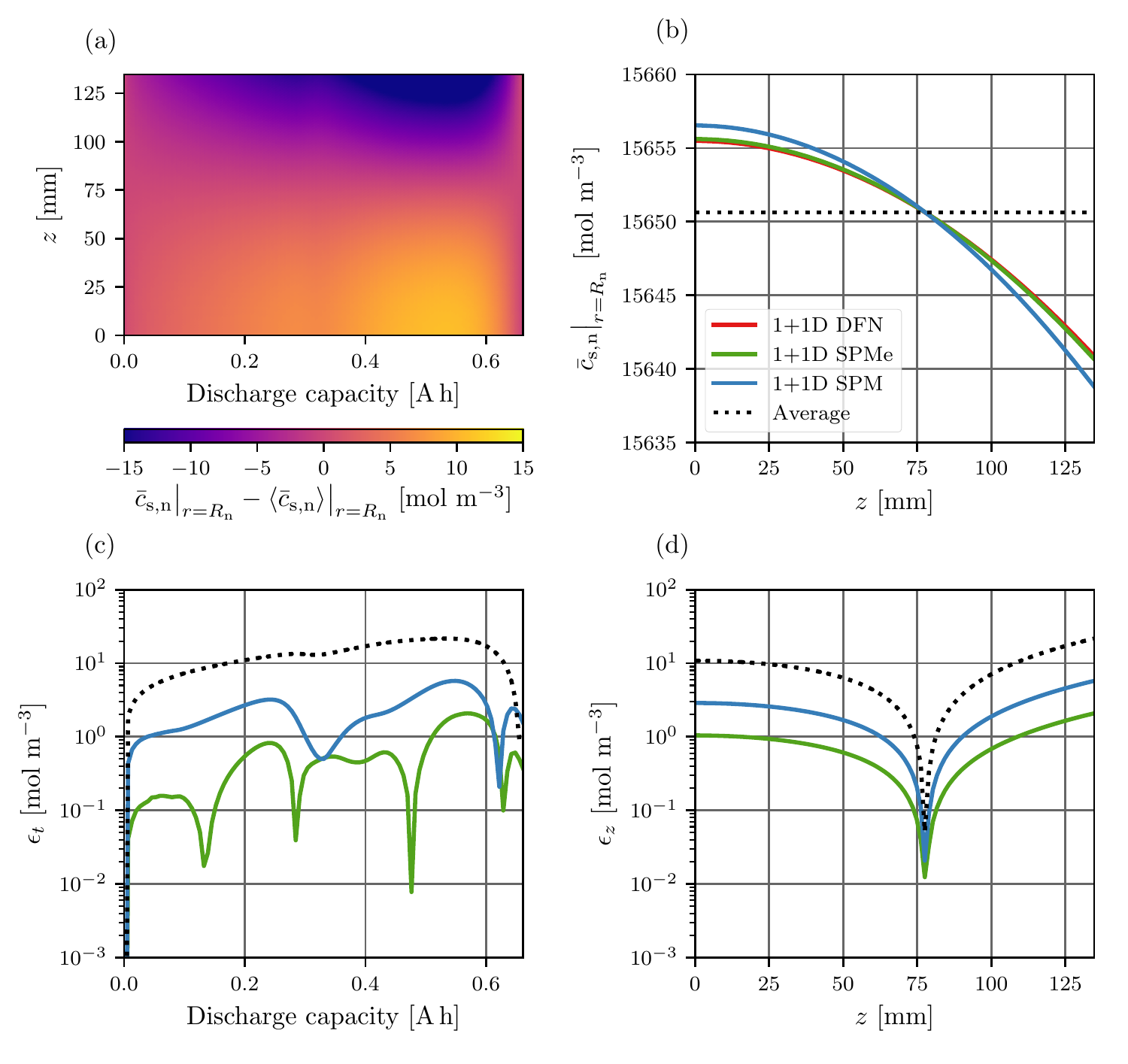}
    \caption{Comparison of the $x$-averaged negative particle surface concentration: (a) Variation in the $x$-averaged negative particle surface concentration predicted by the 1+1D DFN; (b) Snapshot at 0.17[A.h]; (c) Maximum (over $z$) absolute error at each time in the discharge; (d) Maximum (over the entire discharge) absolute error at each $z$ location.}
    \label{fig:particle-concentrations}
\end{figure}

In Figure~\ref{fig:particle-concentrations}(a), we display the variation in $z$ of the surface concentration of the through-cell averaged ($x$-averaged) negative particle predicted by the 1+1D DFN throughout the full discharge. Initially the surface concentration is uniform in $z$. However, as the discharge proceeds, the upper portion of the cell near the tabs becomes more depleted than the lower portion of the cell. This is due to current preferentially travelling through the upper portion of the cell as the path of least resistance. When the upper portion of the cell is almost fully depleted, the resistances associated with discharging the upper portion of the cell increase (i.e. it becomes harder to remove more lithium from the negative electrode and harder to insert lithium into the positive electrode) to a sufficient level such that current preferentially travels through the less utilized portions of the cell, thus leading to the final uniform particle surface concentration. The effect of this on the current can be observed in Figure~\ref{fig:current-comparison}(a). 

In Figure~\ref{fig:particle-concentrations}(b), we plot the $x$-averaged negative particle surface concentration at $\SI{0.17}{A.h}$ predicted by the 1+1D DFN, the 1+1D SPM, the 1+1D SPM, and the average concentration which is predicted by the models that do not include the $z$-dependence of the concentrations (i.e. the ``CC'' and 1D models). Given that the variation in surface concentration is on the order of $\SI{15}{mol.m^{-3}}$, employing a $z$-averaged model recovers the surface concentrations to $0.1\%$ accuracy. However, as shown in Figures~\ref{fig:particle-concentrations}(c) and (d), an order of magnitude decrease in the error can be achieved when using a $z$-resolved model such as the 1+1D SPM or 1+1D SPMe. For situations in which resolving the spatial inhomogenieties (in $z$) in the surface concentration are crucial, a model at least as detailed as the 1+1D SPM should be employed. 

A further comparison of the surface concentration variables is provided in Table~\ref{table:rmse_error_table}. Instead of comparing the $x$-averaged surface concentration, we now compare the surface concentration at every ($x$,$z$)-location in the cell. Here, we observe that the root mean square error (RMSE) in the particle surface concentration is in fact larger for the 1+1D SPM(e) than the 1D DFN and DFNCC. This is because 1+1D SPMe considers a single $x$-averaged particle in the through-cell direction, and in this instance the through-cell variation in the particle surface concentration is more significant than the $z$-direction variation. This highlights the fact that the best combination of through-cell and transverse simplifications depends on which quantities are of interest: for the results here we find that the 1+1D SPMe best recovers the through-cell averaged particle surface concentration, but the 1D DFN better approximates the surface concentration at each ($x$,$z$) location.

\subsection{Comparison of current distribution}
\begin{figure}[htbp]
    \centering
    \includegraphics[width=0.8\textwidth]{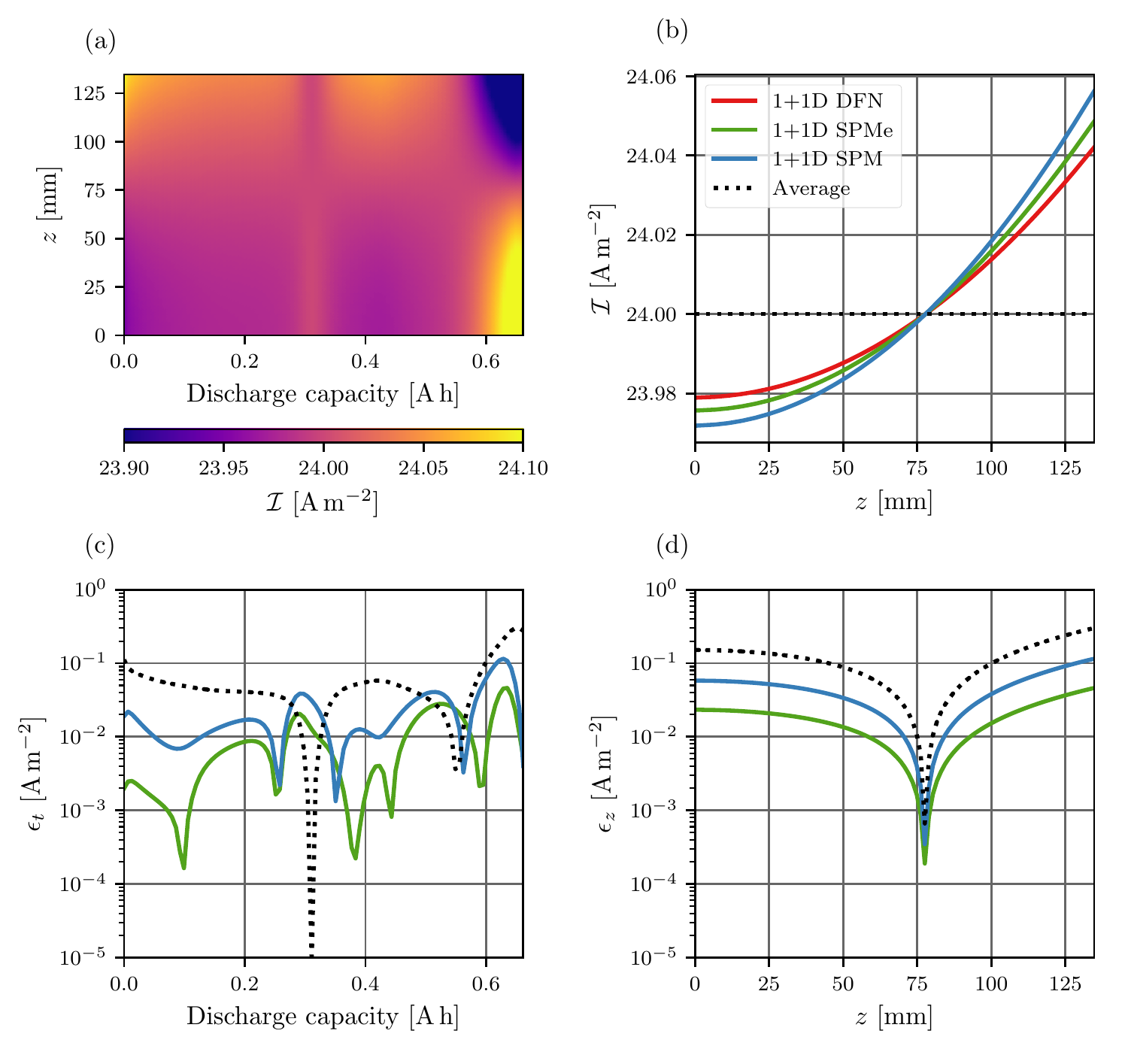}
    \caption{Comparison of the through-cell current: (a) Through-cell current predicted by the 1+1D DFN; (b) Snapshot at 0.17 [A$\,$h]; (c) Maximum (over $z$) absolute error at each time in the discharge; (d) Maximum (over the entire discharge) absolute error at each $z$ location.}
    \label{fig:current-comparison}
\end{figure}

In Figure~\ref{fig:current-comparison}(a), we present the through-cell current density predicted by the 1+1D DFN as a function of $z$ and discharge capacity. As discussed in our description of the particle surface concentration variation, we observe that the current preferentially travels through the top of the cell with the exception of particular times where particle concentrations are such that the through-cell resistance makes it preferential for the current to flow more uniformly through the cell (at $\SI{0.35}{A.h}$) and through the bottom of the cell (at the end of discharge). In Figure~\ref{fig:current-comparison}(b), we show the through-cell current density as a function of $z$ at $\SI{0.17}{A.h}$ through the discharge. We observe that the average current (as predicted by the ``CC'' and 1D models) provides a good first approximation of the through-cell current density which is reasonably uniform (only varying by about $\SI{0.04}{A.m^{-2}}$). However, the 1+1D SPM and 1+1D SPMe both provide an improved estimate of the spatial variation in the through-cell current density, and can give an order of magnitude decrease in the error, as demonstrated in Figures~\ref{fig:current-comparison}(c) and (d). In applications where accurately resolving the current distribution is a key objective and the computational budget is limited, the 1+1D SPMe is the best choice in terms of offering good accuracy at low computational costs. One such application is in studying the effects of degradation, where determining the distribution of current and therefore the overpotentials is crucial for determining rates of degradation. 



\subsection{Comparison of negative potentials}
\begin{figure}[htbp]
    \centering
    \includegraphics[width=0.8\textwidth]{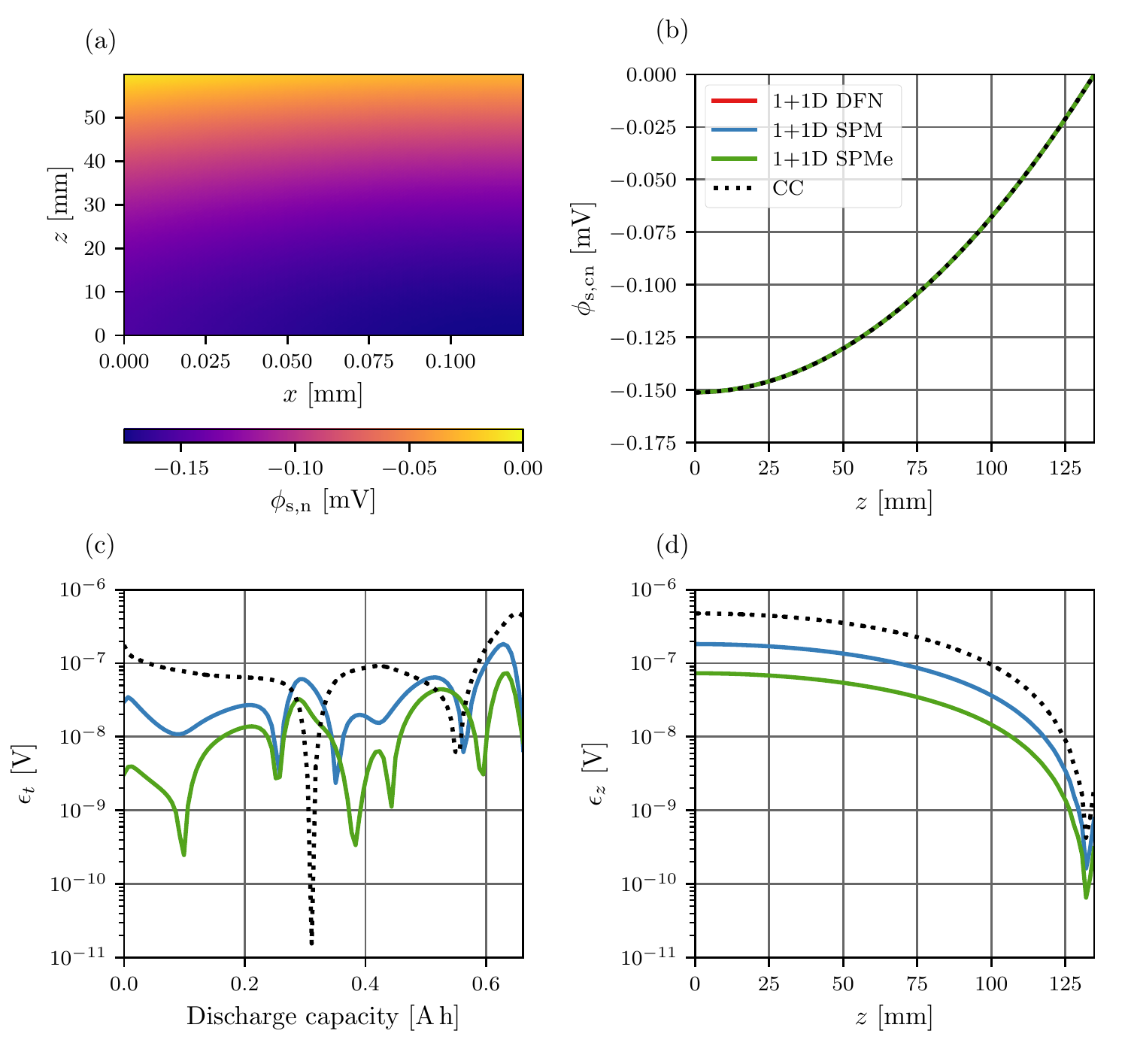}
    \caption{Comparison of negative potentials: (a) Snapshot at 0.17 [A.h] of the negative electrode potential predicted by the 1+1D DFN; (b) Snapshot of the negative current collector potential at 0.17[A.h]; (c) Maximum (over $z$) absolute error in the negative current collector potential at each time in the discharge; (d) Maximum (over the entire discharge) absolute error in the negative current collector potential at each $z$ location. The results of the DFNCC, SPMeCC, and SPMCC are all represented by CC.}
    \label{fig:potential-comparison}
\end{figure}

In Figure~\ref{fig:potential-comparison}(a), we present the potential in the negative electrode at the point $\SI{0.17}{A.h}$ in the discharge (we display this instead of the current collector potential through time because the current collector potential is approximately constant throughout a $\SI{1}{C}$ discharge). In the upper left corner of the electrode, next to the tab, the potential takes values close to the reference value of zero. However, as we move through the cell in the $x$-direction or down the current collector in the $z$-direction we observe a drop in the potential. In particular, a greater potential drop is observed in the current collector direction and so current collector Ohmic losses are of greater importance than through-cell electrode Ohmic losses. In Figures~\ref{fig:potential-comparison}(b), (c), and (d), we compare the negative current collector potential at $\SI{0.17}{A.h}$ as predicted by the 1+1D and ``CC'' models. Note that the 1D models predict that the potential will simply take on the reference value of 0 everywhere in the current collector. We observe excellent agreement between all models. 

\subsection{Summary of Isothermal Comparison}
In this section, we have provided a critical comparison of some of the key outputs predicted by the 1+1D DFN model and the reduced-order models introduced in this paper. In the interest of brevity, we have only presented detailed results for a limited set of variables, but results for other key variables are provided in Table~\ref{table:rmse_error_table} for reference. 

The key result is that in the isothermal case, choosing a reduced-order model that simplifies the $z$-direction behaviour (e.g. the DFNCC) instead of the $x$-direction behaviour (e.g. 1+1D SPMe) provides a better allocation of computational resources. Only in situations where the $z$-direction variation in variables, such as the surface concentrations and through-cell current, are essential should simplifications to the through-cell model be made in favour of simplifications to the transverse model. In such situations, the 1+1D SPMe typically provides a good balance of accuracy and computational cost.

\begin{table}[htb]
	\centering
    \resizebox{\textwidth}{!}{
	\begin{tabular}{c c c | c c | c c c | c c c}
	\toprule
	Variable & Units & Typical values & 1+1D SPMe & 1+1D SPM & DFNCC & SPMeCC & SPMCC & DFN & SPMe & SPM \\
	\midrule
$V$ & $\SI{}{\volt}$ & 3.4 -- 3.8 & $3.25 \times 10^{-3}$ & $2.04 \times 10^{-2}$ & $1.29 \times 10^{-4}$ & $3.29 \times 10^{-3}$ & $2.05 \times 10^{-2}$ & $2.70 \times 10^{-4}$ & $3.33 \times 10^{-3}$ & $2.06 \times 10^{-2}$ \\
$\phi\ts{s, cn}$ & $\SI{}{\volt}$ & $-1.5\times10^{-4}$ -- 0 & $1.44 \times 10^{-8}$ & $3.30 \times 10^{-8}$ & $6.75 \times 10^{-8}$ & $6.75 \times 10^{-8}$ & $6.75 \times 10^{-8}$ & $1.11 \times 10^{-4}$ & $1.11 \times 10^{-4}$ & $1.11 \times 10^{-4}$ \\
$\phi\ts{s,n}$ & $\SI{}{\volt}$ & $-1.7\times10^{-4}$ -- 0 & $2.03 \times 10^{-6}$ & $1.53 \times 10^{-5}$ & $7.79 \times 10^{-8}$ & $2.00 \times 10^{-6}$ & $1.53 \times 10^{-5}$ & $1.11 \times 10^{-4}$ & $1.11 \times 10^{-4}$ & $1.24 \times 10^{-4}$ \\
$\phi\ts{e,k}$ & $\SI{}{\volt}$ & $-0.26$ -- $-0.17$ & $2.03 \times 10^{-3}$ & $1.20 \times 10^{-2}$ & $3.81 \times 10^{-5}$ & $2.03 \times 10^{-3}$ & $1.20 \times 10^{-2}$ & $1.04 \times 10^{-4}$ & $2.07 \times 10^{-3}$ & $1.20 \times 10^{-2}$ \\
$\phi\ts{s,p}$ & $\SI{}{\volt}$ & 3.4 -- 3.8 & $3.25 \times 10^{-3}$ & $2.02 \times 10^{-2}$ & $1.14 \times 10^{-4}$ & $3.25 \times 10^{-3}$ & $2.02 \times 10^{-2}$ & $1.22 \times 10^{-4}$ & $3.28 \times 10^{-3}$ & $2.03 \times 10^{-2}$ \\
$\phi\ts{s,cp}$ & $\SI{}{\volt}$ & 3.4 -- 3.8 & $3.25 \times 10^{-3}$ & $2.04 \times 10^{-2}$ & $7.55 \times 10^{-5}$ & $3.27 \times 10^{-3}$ & $2.04 \times 10^{-2}$ & $1.22 \times 10^{-4}$ & $3.28 \times 10^{-3}$ & $2.05 \times 10^{-2}$ \\
$\mathcal{I}$ & $\SI{}{\ampere.\metre^{-2}}$ & 23.7 -- 24.1 & $6.11 \times 10^{-3}$ & $1.40 \times 10^{-2}$ & $2.87 \times 10^{-2}$ & $2.87 \times 10^{-2}$ & $2.87 \times 10^{-2}$ & $2.87 \times 10^{-2}$ & $2.87 \times 10^{-2}$ & $2.87 \times 10^{-2}$ \\
$c\ts{s,n}\big|_{r=R\ts{n}}$ & $\SI{}{\mol.\metre^{-3}}$ & 4804 -- 19404 & $9.28 \times 10^{2}$ & $9.28 \times 10^{2}$ & $6.77$ & $9.28 \times 10^{2}$ & $9.28 \times 10^{2}$ & $6.77$ & $9.28 \times 10^{2}$ & $9.28 \times 10^{2}$ \\
$c\ts{s,p}\big|_{r=R\ts{p}}$ & $\SI{}{\mol.\metre^{-3}}$ & 31506 -- 48316 & $4.14 \times 10^{2}$ & $4.14 \times 10^{2}$ & $7.33$ & $4.14 \times 10^{2}$ & $4.14 \times 10^{2}$ & $7.33$ & $4.14 \times 10^{2}$ & $4.14 \times 10^{2}$ \\
$c\ts{e,k}$ & $\SI{}{\mol.\metre^{-3}}$ & 800 -- 1200 & $1.08 \times 10^{1}$ & $1.14 \times 10^{2}$ & $1.41 \times 10^{-1}$ & $1.08 \times 10^{1}$ & $1.14 \times 10^{2}$ & $1.41 \times 10^{-1}$ & $1.08 \times 10^{1}$ & $1.14 \times 10^{2}$ \\
$i\ts{e,k}$ & $\SI{}{\ampere.\metre^{-2}}$ & 0.9 -- 24 & $1.65$ & $1.65$ & $2.11 \times 10^{-2}$ & $1.65$ & $1.65$ & $2.11 \times 10^{-2}$ & $1.65$ & $1.65$ \\
$j\ts{k}$ & $\SI{}{\ampere.\metre^{-2}}$ & $-2.3$ -- 2.6 & $3.06 \times 10^{-1}$ & $3.06 \times 10^{-1}$ & $2.54 \times 10^{-3}$ & $3.06 \times 10^{-1}$ & $3.06 \times 10^{-1}$ & $2.54 \times 10^{-3}$ & $3.06 \times 10^{-1}$ & $3.06 \times 10^{-1}$ \\
$\eta\ts{n}$ & $\SI{}{\volt}$ & 0.0014 -- 0.008 & $1.30 \times 10^{-3}$ & $1.34 \times 10^{-3}$ & $1.01 \times 10^{-5}$ & $1.30 \times 10^{-3}$ & $1.34 \times 10^{-3}$ & $1.01 \times 10^{-5}$ & $1.30 \times 10^{-3}$ & $1.34 \times 10^{-3}$ \\
$\eta\ts{p}$ & $\SI{}{\volt}$ & $-0.12$ -- $-0.06$ & $4.20 \times 10^{-3}$ & $4.80 \times 10^{-3}$ & $7.14 \times 10^{-5}$ & $4.20 \times 10^{-3}$ & $4.80 \times 10^{-3}$ & $7.14 \times 10^{-5}$ & $4.20 \times 10^{-3}$ & $4.80 \times 10^{-3}$ \\
    \bottomrule
	\end{tabular}
	} 
    \caption{RMSE of key model variables in each of the reduced-order models vs. the 1+1D DFN for a $\SI{1}{C}$ constant current discharge.} 
    \label{table:rmse_error_table}
\end{table}

\subsection{Critical Comparison of 1+1D Thermal Models}
We now compare the thermal versions of the models under the conditions of a $\SI{3}{C}$ discharge and tab cooling. Tab cooling is simulated by applying
\begin{equation}\label{eqn:tab-cool}
    \lambda \nabla T\cdot \boldsymbol{n} = -h\ts{tab} (T-T_\infty)
\end{equation}
on the tabs and
\begin{equation}
    \lambda \nabla T\cdot \boldsymbol{n} = 0
\end{equation}
on all other boundaries. For the purposes of this comparison, both tabs are placed at the top of the cell and the value of $h\ts{tab}$ is set to $\SI{1000}{W.m^{-2}.K^{-1}}$ so that a temperature variation on the order of a few degrees Kelvin is observed across the cell; this is in accordance with experimental results \cite{hunt2016}. This approach was taken as a simple way to induce a variation in the temperature in the $z$-direction that is similar to the variation seen in experiments. However, a proper treatment of tab cooling would involve a more complete model of the tab which has been shown to be a major heat transfer bottle neck \cite{hunt2016}. 

\begin{figure}[htbp]
    \centering
    \includegraphics[width=0.8\textwidth]{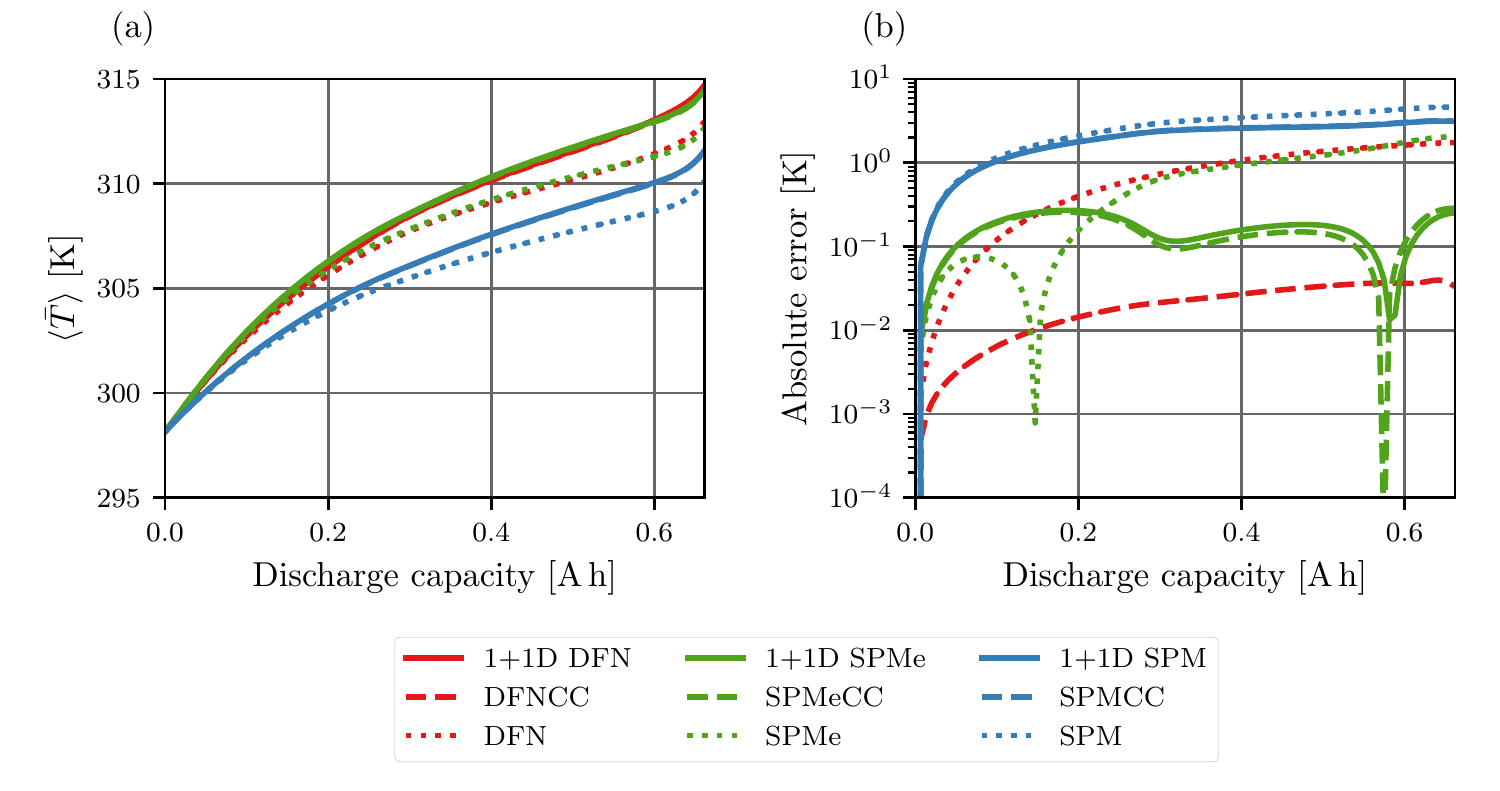}
    \caption{Comparison of the volume-averaged cell temperature: (a) predicted volume-averaged cell temperature; (b) absolute error between reduced model and 1+1D DFN at each time during discharge.}
    \label{fig:vol-av-temp}
\end{figure}

Figure~\ref{fig:vol-av-temp} we show the volume-averaged temperature predicted by each model. Owing to the effects of tab cooling, it is important to account for the current collectors in order to accurately predict the temperature. However, a large contribution to the error comes from ignoring electrolyte effects and assuming uniform (in $x$) electrode potentials, as evidenced by the poor performance of all of the models that use the SPM. For the parameters in Tables~\ref{table:dimensional_parameter_values} and \ref{table:dimensional_coefficients} it is better to include a more complex model of the electrochemistry with a simpler model of the transverse behaviour in order to accurately predict the temperature. For instance, it is a better use of computational resources to choose the DFNCC rather than the 1+1D SPMe. 


The DFN, SPMe and SPM employ a lumped thermal model with cooling proportional to the difference between the average temperature and the ambient temperature. Since the actual temperature at the tab is lower than the average temperature, these models overpredict the cooling rate, giving a lower volume-averaged cell temperature, as observed in Figure~\ref{fig:vol-av-temp}.

\begin{figure}[htbp]
    \centering
    \includegraphics[width=0.8\textwidth]{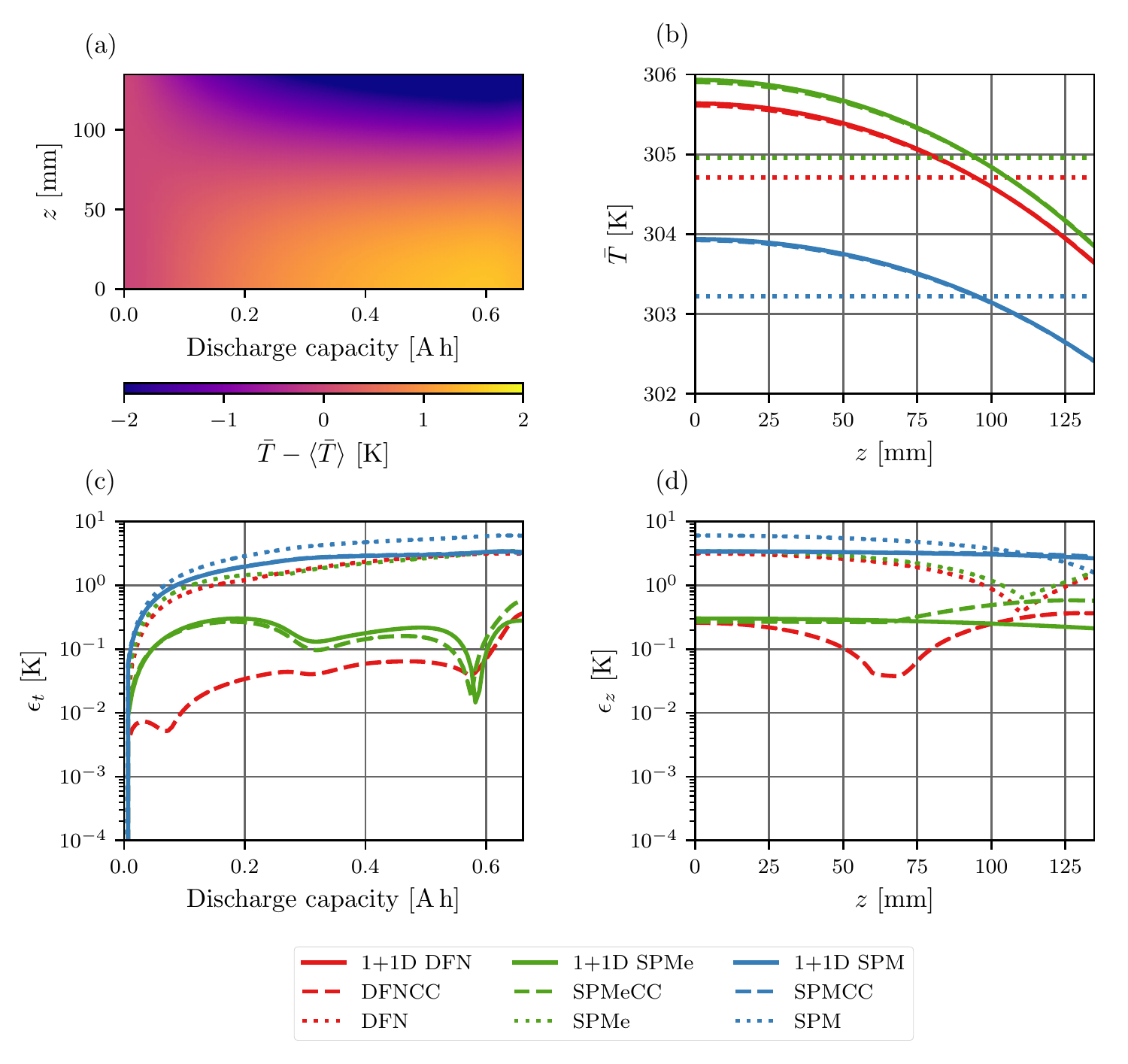}
    \caption{Comparison of temperature profiles: (a) $x$-averaged cell temperature variation through time predicted by the 1+1D DFN; (b) $x$-averaged cell temperature at 0.17 [A$\,$h]; (c) Maximum (over $z$) absolute error at each time in the discharge; (d) Maximum (over the entire discharge) absolute error at each $z$ location.}
    \label{fig:y-z-temp-prof}
\end{figure}

Figure~\ref{fig:y-z-temp-prof} shows the temperature as a function of space and time throughout the discharge. Again we observe that the DFNCC best predicts the temperature profile of the 1+1D DFN, and the greatest error is introduced by using the SPM. Note that the error for the 1+1D and ``CC'' models is similar, backing up the suggestion to use a simpler transverse model (that still retains some $z$ dependence).


\begin{figure}[htbp]
    \centering
    \includegraphics[width=0.8\textwidth]{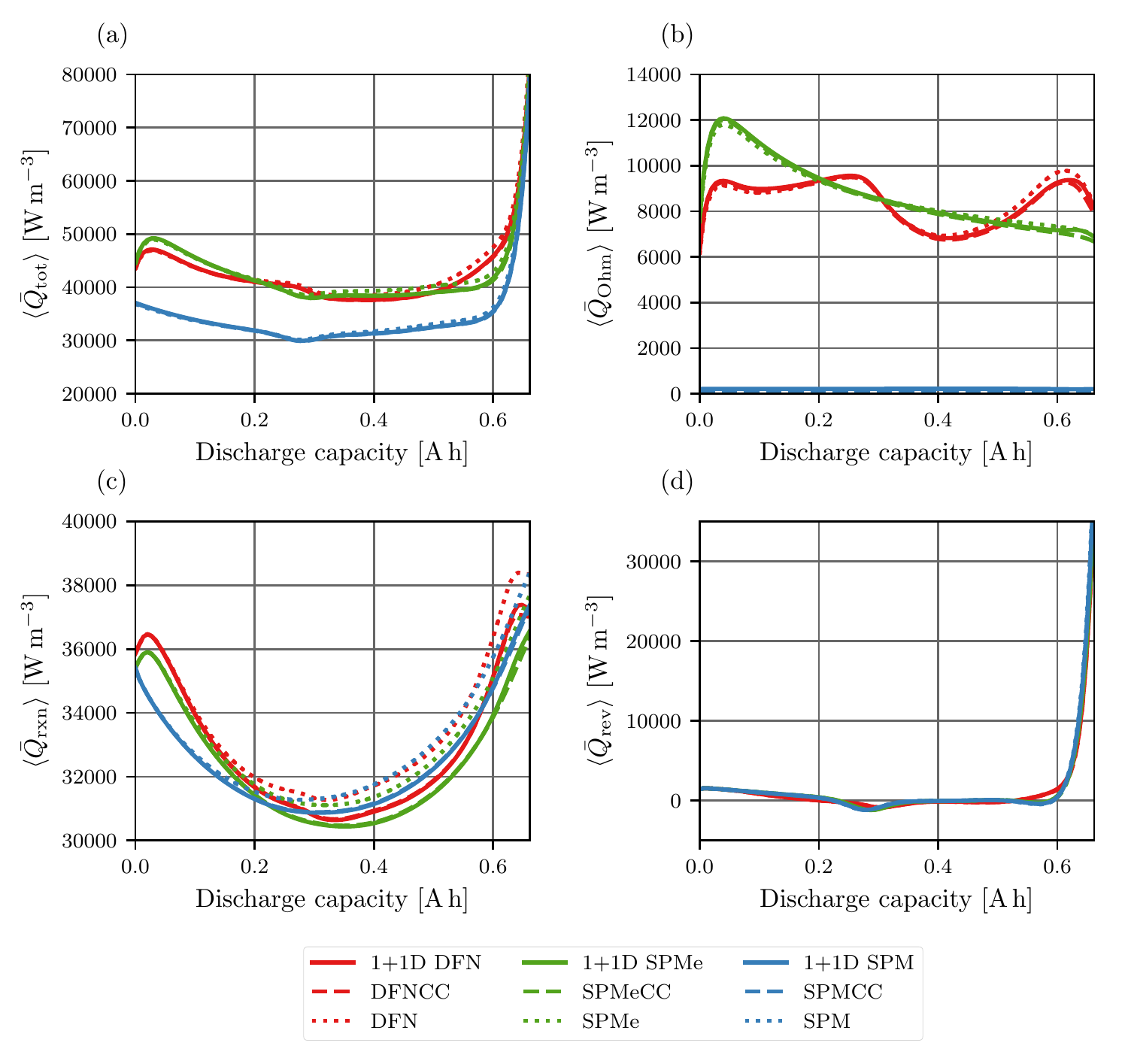}
    \caption{Comparison of volume-averaged cell heating: (a) total heating; (b) Ohmic heating; (c) reaction (irreversible) heating; (d) irreversible heating.}
    \label{fig:vol-av-heating}
\end{figure}

In Figure~\ref{fig:vol-av-heating}, we break down the volume-averaged heating into its individual components to help diagnose where errors in the various reduced models arise. We observe that across all forms of heating, the DFNCC  is almost indistinguishable from the results of the 1+1D DFN. Whilst one might initially think that the 1D DFN should produce the same irreversible and reversible reaction heating as the DFNCC, this is not not the case because the temperature predicted by the 1D DFN is lower, as mentioned in the discussion of Figure~\ref{fig:vol-av-temp}. The temperature dependence of the electrochemical reactions then means that the 1D DFN actually overpredicts the reaction heating (an effect which increases as the temperatures diverges further through the discharge). The SPMeCC and 2+1D SPMe both capture the general behaviour of the Ohmic and irreversible reaction heating, however they fail to capture fluctuations in this general behaviour, particularly in the Ohmic heating. This is a result of failing to capture though-cell variations in the reaction overpotentials, as well as the more detailed variations in the electrolyte potentials. Despite this, both the SPMeCC and 1+1D SPMe perform reasonably well at recovering the total heating. The main failing of the SPM and 1+1D SPM is that neither accounts for any though-cell Ohmic heating, with the 1+1D SPM only accounting for current collector Ohmic heating. As a result, these models significantly under predict the cell heating. Therefore it is recommended that a more complicated model than the SPM is used for the through-cell model for thermal studies, unless the model parameters vary significantly from those in this study. 

\begin{figure}[htbp]
    \centering
    \includegraphics[width=0.8\textwidth]{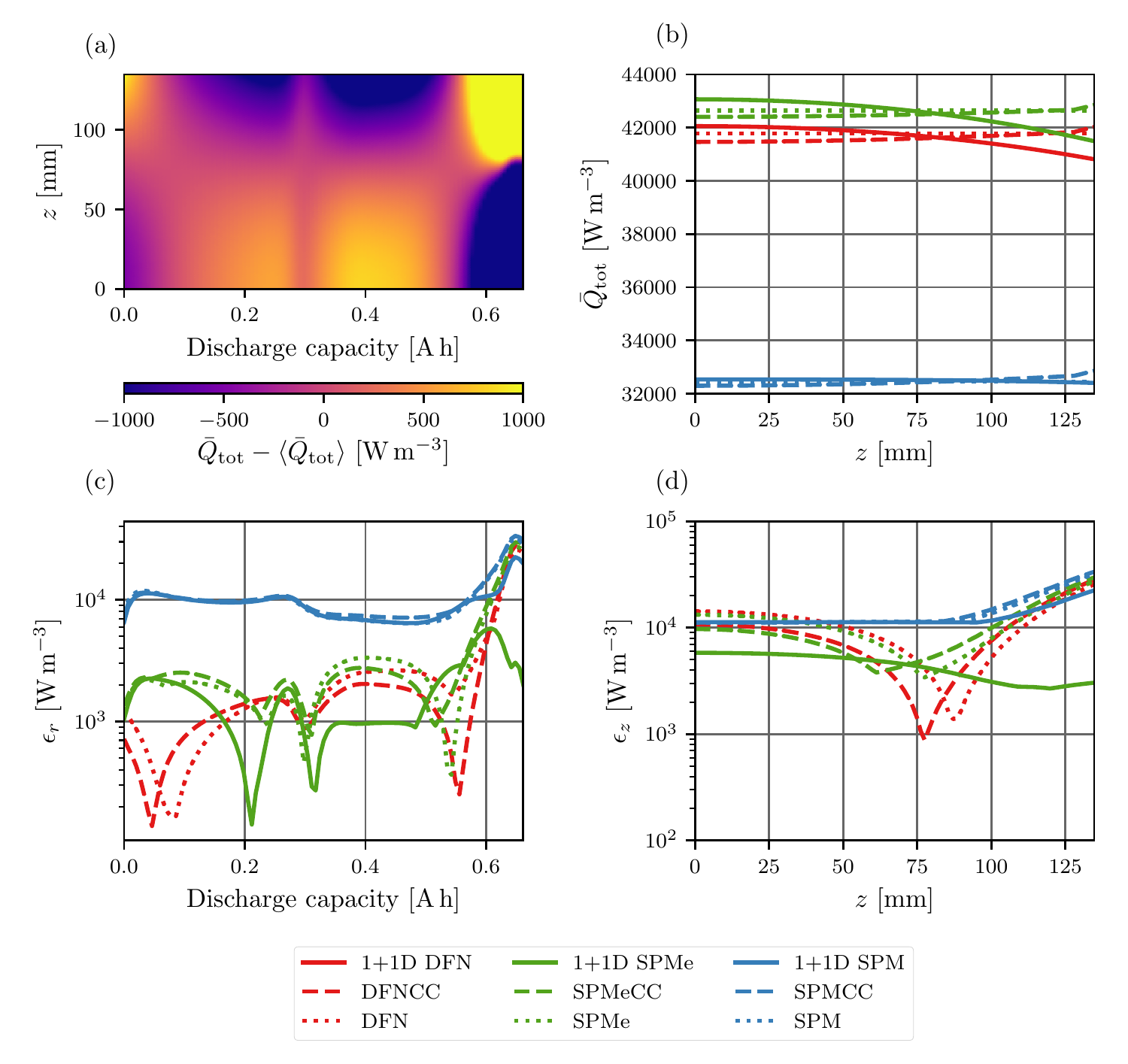}
    \caption{Comparison of $x$-averaged total heating, $\Xav{Q}\ts{tot}$: (a) The variation in the heat generation profile throughout the discharge predicted by the 1+1D DFN; (b) $x$-averaged heat generation at 0.17[A.h]; (c) Maximum (over $z$) absolute error at each time in the discharge; (d) Maximum (over the entire discharge) absolute error at each $z$ location.}
    \label{fig:y-z-heat-prof}
\end{figure}

\begin{figure}[htbp]
    \centering
    \includegraphics[width=0.8\textwidth]{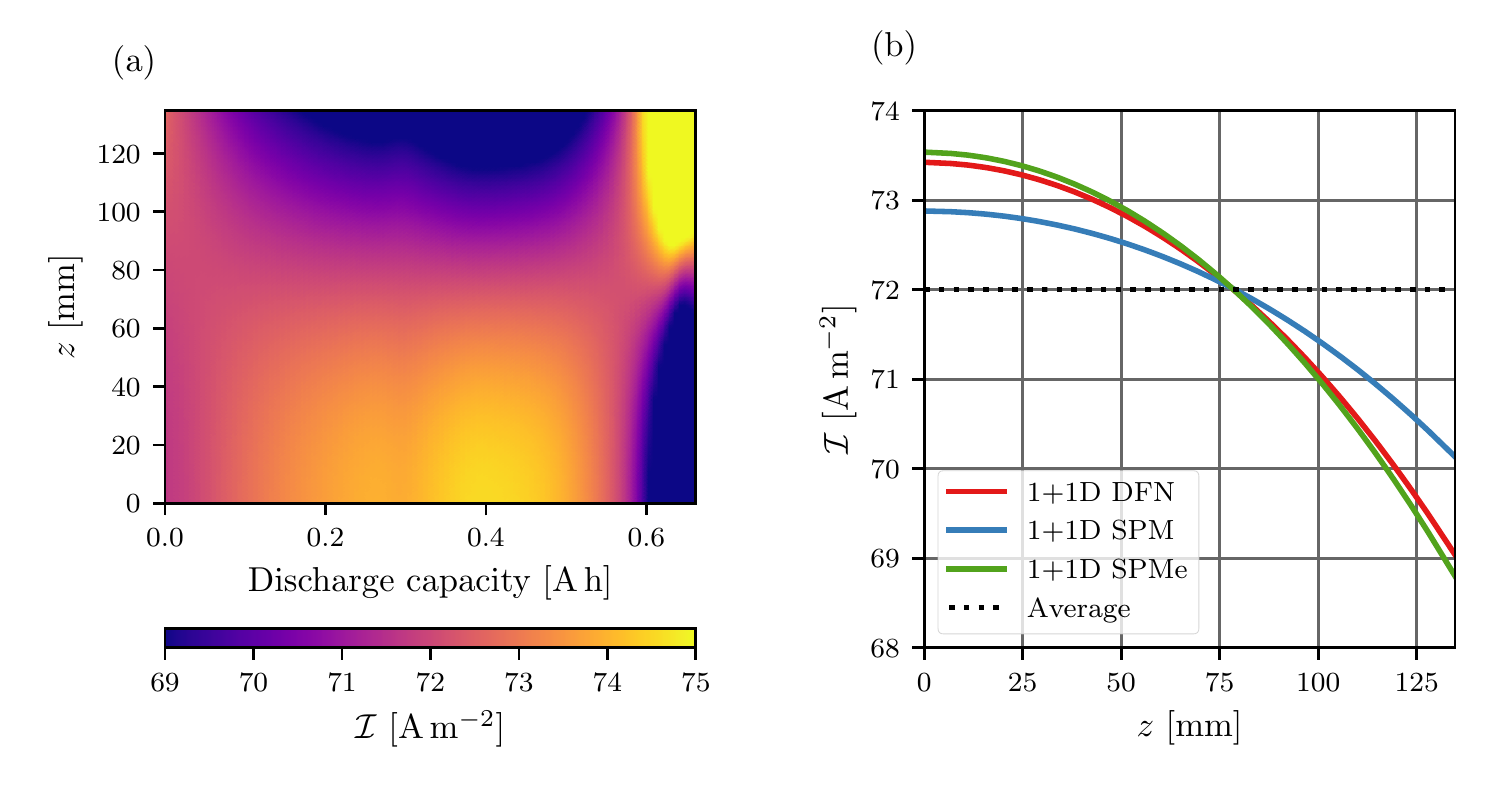}
    \caption{Current distribution during discharge of tab-cooled cell.}
    \label{fig:thermal-current}
\end{figure}

In Figure~\ref{fig:y-z-heat-prof} (a), we present the total heating predicted by the 1+1D DFN as a function of $z$ and the discharge time. Initially the cell heats near the top, but as the discharge proceeds heating mainly occurs near the bottom of the cell. This may seem in contrast to what we would expect given the isothermal current profiles in Figure~\ref{fig:current-comparison}. However, as shown in Figure~\ref{fig:thermal-current}, the current profile can in fact be very different for the tab cooling scenario considered here. We now see that the current now preferentially travels through the bottom of the cell for most of the discharge, and the increased current leads to increased heating. This is an effect of temperature dependence of the parameters in the through-cell models: higher temperature leads to lower resistance. In Figures~\ref{fig:y-z-temp-prof} (b), (c), and (d), we compare the heat generation predicted by each reduced order model with that of the 1+1D DFN. We observe that all reduced models give rise to around a 10$\%$ error in the total heating. 

\subsection{Drive cycle comparison}
\begin{figure}[htbp]
    \centering
    \includegraphics[width=0.8\textwidth]{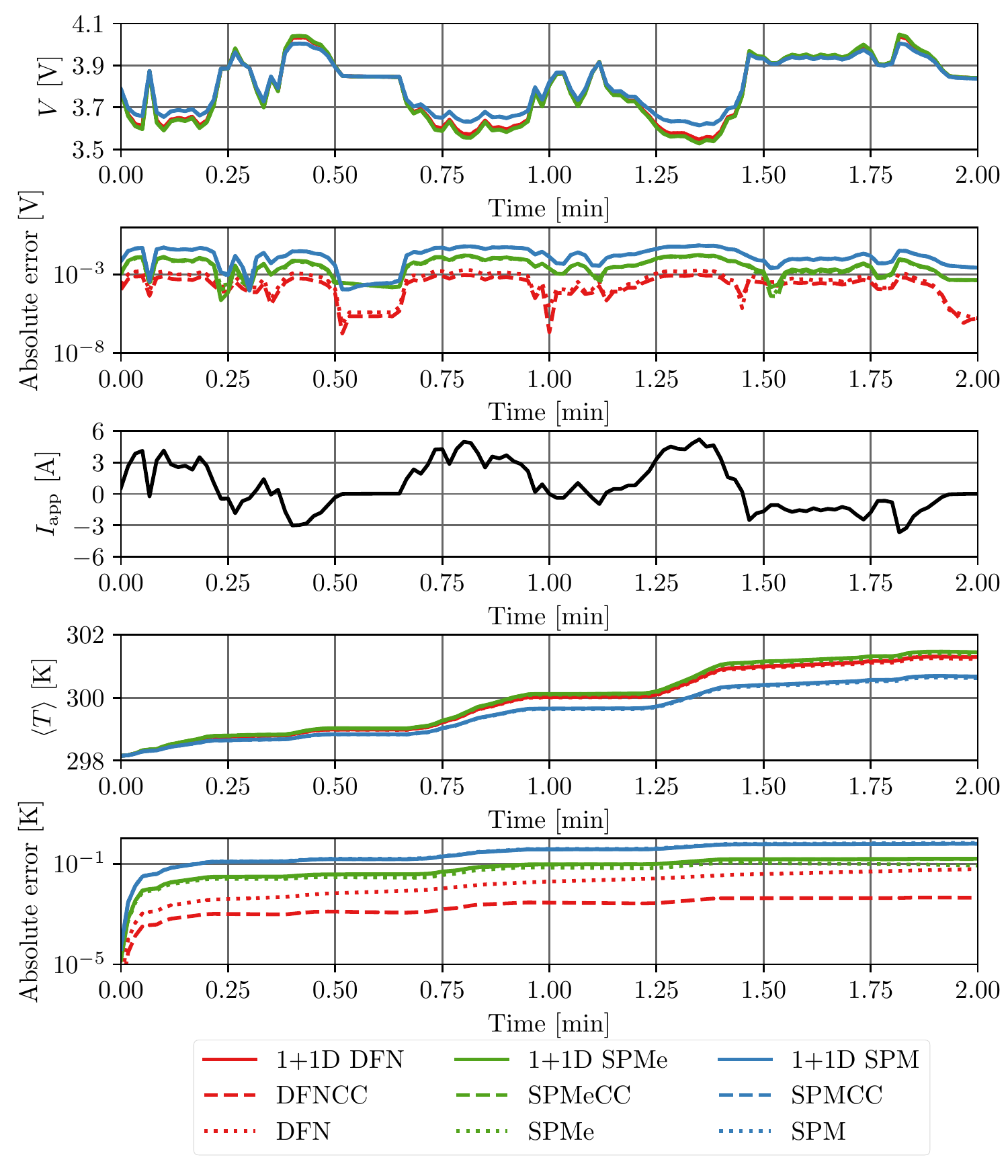}
    \caption{Comparison of model performance on US06 drive cycle. Solve times were: 1+1D DFN 154s, 1+1D SPMe 83s, 1+1D SPM 11s, DFN(CC) 25s, SPMe(CC) 1s, SPM(CC) 0.23s}
    \label{fig:drive_cycle}
\end{figure}

The constant current discharges that we have investigated throughout this paper are useful for comparing models. However, they are not fully representative of a realistic usage scenario. To give an example of how the models perform in more realistic conditions, we compare the performance of each model on a portion of the US06 drive cycle. Here, we just consider the measured terminal voltage and the average temperature of the cell, as shown in Figure~\ref{fig:drive_cycle}. We observe that similar to our previous results, the main errors are introduced by making simplifications to  the through-cell electrochemical model. Therefore to most accurately capture the temperature rises, computational effort should be placed upon using a more detailed through-cell model like the DFN with a CC or 0D transverse model, rather than in using a detailed 1+1D  transverse model and a simplified through-cell model like the SPM.  

\section{Comparison Summary}\label{sec:comparison-summary}
In the previous sections, we have provided a detailed comparison of the nine models in Figure~\ref{fig:model_schematic}. In Table~\ref{tab:summary-table}, we have aimed to condense this information into a concise and clear format. Whilst our recommendations presented in Table~\ref{tab:summary-table} are informed by the quantitative performance of the models in each of the comparisons in the previous sections, they are to some degree qualitative in nature. 

The first two columns in the table correspond to the solve time and the number of states required to solve that model. The solve times should only be considered relative to one another, as one could achieve speed ups by employing different numerical methods or hardware; they are representative of the time complexitity of each model. Similarly, the number of states should only be considered as representative of the spatial complexity of each model as different discretization methods may lead to a different number of states. Solve times in green are on the order of 10ms, orange on the order of 100ms, and red on the order of 1000ms, so that a green model is approximately 100 times faster than a red model. We similarly colour the states of each model so that a red model requires around 10 times more memory than a orange model and around 100 times more memory than a green model. 

The second set of columns summarises the ability of each model to accurately predict five key output variables during a $\SI{1}{C}$ discharge: the terminal voltage, $V$; the current distribution, $\mathcal{I}$; the negative current collector potential, $\mathcal{\phi}\ts{s,cn}$; the $x$-averaged negative particle surface concentration, $\bar{c}\ts{s,n}$; and the average cell temperature, $\XYZav{T}$. We adopt a traffic light system for each of these variables as described in Table~\ref{tab:traffic-system-key}. The system is designed to divide the predictions of each variable into three categories, where green is most accurate and red is least accurate. This is done with reference to the results presented throughout this paper.  Generally, moving from one color to the next represents an order of magnitude difference in error, but this is not always the most appropriate division and we refer the reader to Table~\ref{tab:traffic-system-key} for precise details. To be clear, red does not indicate that the model should not be used, but rather that this model performs poorly relative to the other models. Since we refer to the 1+1D DFN model for calculation of errors in models, the 1+1D DFN is always coloured green. 

The third set of columns considers the ability of each model to predict the terminal voltage under low current, medium current, and high current conditions. We again adopt a traffic light system where a maximum absolute error of $<\SI{1e3}{\volt}$ is coloured green, $<\SI{1e-1}{V}$ is coloured orange, and $<\SI{1}{V}$ is colored red. The data used for these columns summarises that in Figure~\ref{fig:terminal-voltage-comparison}(c). Again, since we refer to the 1+1D DFN model for calculation of errors in models, the 1+1D DFN is always coloured green. 

\begin{table}[h]
    \centering
	\resizebox{\textwidth}{!}{%
    \begin{tabular}{l | c c | c c c | c | c c c}\toprule
         \multirow{2}{4em}{Model} & \multirow{2}{7em}{Solve time [ms]}  & \multirow{2}{3em}{States} & \multicolumn{3}{c|}{$\SI{1}{C}$} & $\SI{3}{C}$ &  \multicolumn{3}{c}{$V$} \\
         & & & $\mathcal{I}$ & $\phi\ts{s,cn}$ & $\bar{c}\ts{s,n}$ & $\XYZav{T}$ & $I\ts{app}<\SI{1}{C}$ & $\SI{1}{C}<I\ts{app}<\SI{4}{C}$& $I\ts{app}>4C$ \\ \midrule
         1+1D DFN & \bad{8377} & \bad{49561} & \cmark & \cmark & \cmark & \cmark & \cmark & \cmark & \best \\ 
         1+1D SPMe & \bad{1519} & \ok{3961} & \cmark & \cmark & \cmark & \okmark & \okmark & \okmark & \xmark \\ 
         1+1D SPM & \ok{102} & \ok{1261} & \okmark & \okmark & \okmark & \xmark & \okmark & \xmark & \xmark \\
         DFNCC & \ok{248} & \ok{1651} & \xmark & \okmark & \xmark & \cmark & \cmark & \cmark & \cmark \\ 
         SPMeCC & \good{10} & \good{131} & \xmark & \okmark & \xmark & \okmark  & \okmark & \okmark & \xmark \\ 
         SPMCC & \good{5} & \good{41} & \xmark & \okmark & \xmark & \xmark & \okmark & \xmark & \xmark \\ 
         DFN & \ok{248} & \ok{1651} & \xmark & \xmark & \xmark & \xmark & \cmark & \cmark  & \cmark \\ 
         SPMe & \good{10} & \good{131} & \xmark  & \xmark & \xmark & \xmark  & \okmark & \okmark & \xmark  \\ 
         SPM & \good{5} & \good{41}  & \xmark & \xmark & \xmark & \xmark & \okmark & \xmark & \xmark \\ \bottomrule
    \end{tabular}
    }
    \caption{Qualitative evaluation of model. In the variables, we employed the traffic light system described in Table~\ref{tab:traffic-system-key}. For the final set of columns (current dependence), we make use of the results in Figure~\ref{fig:terminal-voltage-comparison}(c) with the following traffic light system for the maximum absolute errors: $<\SI{1e-3}{\volt}$ (green), $<\SI{1e-1}{V}$ (orange), $<\SI{1}{V}$ (red).}
    \label{tab:summary-table}
\end{table}

\begin{table}[h]
    \centering
    \begin{tabular}{l | c c c c c }\toprule
         & $V$ [$\SI{}{\volt}$]& $\mathcal{I}$ [$\SI{}{\ampere.\metre^{-2}}$] & $\phi\ts{s,cn}$ [$\SI{}{\volt}$]& $\bar{c}\ts{s,n}$ [$\SI{}{\mol.\metre^{-3}}$]& $\XYZav{T}$ [$\SI{}{\kelvin}$] \\ \midrule
         \cmark & $<10^{-3}$ & $<5\times10^{-2}$ & $<10^{-7}$ & $<5$ & $<10^{-1}$ \\ 
         \okmark & $<10^{-1}$ & $<10^{-1}$ & $<10^{-6}$ & $<10$ & $<1$ \\ 
         \xmark & $<1$ & $<1$ & $<10^{-3}$ & $<100$ & $<10$ \\ \bottomrule
    \end{tabular}
    \caption{Traffic light system key. The numbers refer to ranges in the absolute errors obtained for a $\SI{1}{C}$ discharge ($\SI{3}{C}$ for temperature). For the variables, $V$, $\mathcal{I}$, $\phi\ts{s,cn}$, and $\bar{c}\ts{s,n}$ these errors refer to the isothermal case. For the variable $T$, the errors refer to the thermal case.}
    \label{tab:traffic-system-key}
\end{table}

We now interpret the summary presented in Table~\ref{tab:summary-table}. Firstly, by employing the 1+1D SPMe instead of the 1+1D DFN, we can achieve similar performance in terms of predicting the current, current collector potential, and y-z concentration variation while reducing memory requirements by an order of magnitude. This is achieved at the expense of a slight reduction in accuracy of the terminal voltage prediction and average cell temperature. The model is most appropriate in the low to medium C-rate range. Employing the 1+1D SPM sees a similar reduction in memory requirements, but with an additional reduction in solve time by an order of magnitude. However, this is achieved at the expense of less accurate predictions of all of the variables as well as being limited to low C-rates. 

The DFNCC offers a reduction in both solve time and memory requirements by an order of magnitude without loss of accuracy in both the terminal voltage and average cell temperature. Further, there is only a small reduction in accuracy of the current collector potential. However, it achieves this reduction in solve time and memory requirements at the expense of accuracy in the estimates of the through-cell current density, and $x$-averaged negative particle concentration. The DFNCC remains applicable across the range of C-rates we investigated. The SPMeCC drops the solve time and memory requirements by a further order of magnitude relative to the DFNCC. To achieve this a small amount of accuracy is sacrificed in the terminal voltage and average temperature estimates. Additionally, the SPMeCC is limited to low and medium C-rates. The solve time and memory requirements of the SPMCC are similar to that of the SPMeCC, but the predictions of the voltage and average temperature are less accurate. Further, the SPMCC is limited to low C-rates. 

The DFN, SPMe, and SPM all offer orders of magnitude reductions in solve time and memory requirements compared to the models that account for current collector effects, but all do so at the expense of accuracy in the predictions of through-cell current density, current collector potential, $x$-averaged concentration, and average temperature. However, the DFN can accurately predict the terminal voltage, and performs well across the full range of C-rates we considered. The SPMe performed moderately at recovering the terminal voltage, and moderately well at low to medium C-rates. The SPM was the worst performing model across all variables, and is best applied at low C-rates.

Table~\ref{tab:summary-table} highlights the trade-off that must be made between computational complexity and accuracy. However, we see that through the appropriate choice of a model, one can choose where accuracy is sacrificed in favour of reduced complexity. For example, in a study of spatially dependant degradation within a lithium-ion pouch cell, it makes more sense to employ the 1+1D SPMe or the 1+1D SPM instead of the DFN or DFNCC, because they retain greater accuracy in the y-z dependent variables for a similar computational budget. Alternatively, in pack or module level simulations where one is only interested in the average temperature and voltage outputs of a cell, the DFNCC or SPMeCC are more appropriate. Further, where computational constraints are really strict such as in model-based control, simple models such as the SPM and SPMe are most appropriate (the SPMCC and the SPMeCC could also be used in a limited form). In applying these models it is important to understand their limitations so that they can be used appropriately. As a general rule, if integrated quantities, such as the terminal voltage, are of most importance then the most detailed through-cell model that the budget can afford should be used. On the other hand, if capturing variations in the transverse directions is important then a 1+1D model is an appropriate choice.

\section{Conclusions}\label{sec:Conclusions}
In this paper we have presented a hierarchy of reduced-order models of a lithium-ion pouch cell of varying fidelity and computational complexity. Simplifications to the transverse and through-cell behaviour, derived via asymptotic analysis \cite{Marquis2019, sulzer_part_I}, have been combined in a consistent way, so that they can be used in any combination to suit the particular application. In particular, we note that the transverse and through-cell model can be selected independently, and the framework allows for additional effects to be included in a straightforward manner. The asymptotic approach allows for \emph{a priori} estimates of the modelling error to be made so that the appropriateness of applying a  particular reduced-order model can be assessed prior to implementation. In addition, we have provided a numerical comparison of the reduced-order models and the full 1+1D DFN model.

Through a series of comparisons it was demonstrated that the choice of reduced-order model depends on the variables of interest for a particular application. For instance, in many control or systems-level applications, one is only concerned with obtaining integrated quantities, such as the terminal voltage and volume-averaged cell temperature. In such cases it is best to select the highest fidelity through-cell models the computational budget allows, combined with a simpler transverse model (e.g. the SPMeCC). However, if distributed (in $y$-$z$) quantities are of interest, such as in trying to model non-uniform degradation, then it is necessary to choose a more complicated transverse model (e.g. the 1+1D SPMe). 

\clearpage
\appendix

\newpage
\section{Parameter Values}
\begin{table}[htb]
	\centering
	\resizebox{\textwidth}{!}{%
	\begin{tabular}{c c p{7cm} c c c c c}
	\toprule
     Parameter & Units & Description & cn & n & s & p & cp \\
    \midrule
     $L\ts{k}$ & $\SI{}{\micro\metre}$ & Region thickness & 25 & 100 & 25 & 100 & 25 \\
     $L\ts{tab,ck}$ & $\SI{}{\milli\metre}$ & Tab width & - & 40 & - & 40 & - \\
    $c\ts{e,typ}$ & $\SI{}{\mol.\metre^{-3}}$ & Typical lithium concentration in electrolyte & - & $1\times10^3$ & $1\times10^3$ & $1\times10^3$ & - \\
    $D\ts{e,typ}$ & $\SI{}{\metre^{2}.\second^{-1}}$ & Typical electrolyte diffusivity  &- & $5.34\times10^{-10}$ & $5.34\times10^{-10}$ & $5.34\times10^{-10}$ & - \\
    $\epsilon\ts{k}$ &-  & Electrolyte volume fraction & - & $0.3$ & 1 & $0.3$ & - \\
    $c\ts{s,k,max}$ & $\SI{}{\mol.\metre^{-3}}$ & Maximum lithium concentration in solid  & - & $2.498\times 10^4$ & - & $5.122\times 10^4$ & - \\
     $\sigma\ts{k}$ & $\SI{}{\ohm^{-1}.\metre^{-1}}$ & Solid conductivity  & $5.96\times10^{7}$ & 100 & - & 10 & $3.55\times10^7$ \\
     $D\ts{s,k,typ}$ & $\SI{}{\metre^{2}.\second^{-1}}$ & Typical solid diffusivity  & - &  $3.9\times10^{-14}$ & - &  $1\times10^{-13}$ & - \\
     $R\ts{k}$ & $\SI{}{\micro\metre}$ & Particle radius  & - & 10 & - & 10 & -  \\
     $a\ts{k}$ & $\SI{}{\micro\metre^{-1}}$ & Electrode surface area density  & - & 0.18 & - & 0.15 & - \\
     $m\ts{k,typ}$ & $\SI{}{\ampere.\metre^{-2}.(\metre^3.\mol^{-1})^{1.5}}$ & Typical reaction rate  & - & $2\times10^{-5}$ & - & $6\times10^{-7}$ & - \\
     $\rho\ts{k}$ & $\SI{}{\kilogram.\metre^{-3}}$ & Density  & 8954 & 1657 & 397 & 3262 & 2707 \\
     $c\ts{p,k}$ & $\SI{}{\joule.\kilogram^{-1}.\kelvin^{-1}}$ & Specific heat capacity  & 385 & 700 & 700 & 700 & 897 \\
     $\lambda\ts{k}$ & $\SI{}{\watt.\metre^{-1}.\kelvin^{-1}}$ & Thermal conductivity  & 401 & 1.7 & 0.16 & 2.1 & 237 \\
   \midrule
   $E_{m\ts{k}}$ & $\SI{}{\joule.\mol^{-1}}$ & Activation energy for reaction rate & - & $3.748 \times 10^4$ & - & $3.957 \times 10^4$ & - \\
    $E_{D\ts{e}}$ & $\SI{}{\joule.\mol^{-1}}$ & Activation energy for electrolyte diffusivity &  &  & $3.704 \times 10^4$ &  &  \\
    $E_{\kappa\ts{e}}$ & $\SI{}{\joule.\mol^{-1}}$ & Activation energy for electrolyte conductivity &  &  & $3.470 \times 10^4$ &  &  \\
    \midrule
    $c_{k,0}$ & $\SI{}{\mol.\metre^{-3}}$ & Initial lithium concentration in solid  & - & $1.999\times10^4$ & - & $3.073\times10^4$ & - \\
    $T_0$ & $\SI{}{\kelvin}$ & Initial temperature & & & 298.15 & & \\
    \midrule
     $F$ & $\SI{}{\coulomb.\mol^{-1}}$ & Faraday's constant  & & & 96487 & & \\
     $R_g$ & $\SI{}{\joule.\mol^{-1}.\kelvin^{-1}}$ & Universal gas constant  &&& 8.314 && \\
     $T_\infty$  & $\SI{}{\kelvin}$ & Reference temperature &&& 298.15 && \\
     $b$ & - & Bruggeman coefficient  &&& 1.5 &&  \\
     $t^+$ & - & Transference number &&& 0.4 && \\
     $L_x$ & $\SI{}{\micro\metre}$ & Cell thickness  &&& 225 && \\
     $L_y$ & $\SI{}{\milli\metre}$ & Cell width  &&& 207 && \\
     $L_z$ & $\SI{}{\milli\metre}$ & Cell height  &&& 137 && \\
    $I\ts{app}$ & $\SI{}{\ampere}$ & Applied current  &&&  0.681 && \\
    $h$ & $\SI{}{\watt.\metre^{-2}.\kelvin^{-1}}$ & Heat transfer coefficient  &&& 10 && \\
    $\rho\ts{eff}$ & $\SI{}{\joule.\kelvin^{-1}.\metre^{-3}}$ & Lumped effective thermal density & & & $1.812 \times 10^{6}$  & & \\
   $\lambda\ts{eff}$ & $\SI{}{\watt.\metre^{1-}.\kelvin^{-1}}$ & Effective thermal conductivity & & & 59.396  & & \\    
    \bottomrule
	\end{tabular}}
        \caption{Typical dimensional parameter values taken from \cite{SMouraGithub}. The parameters are for a carbon negative current collector, graphite negative electrode, LiPF$_6$ in EC:DMC electrolyte, LCO positive electrode, and aluminium positive current collector. } \label{table:dimensional_parameter_values}
\end{table}

\begin{table}[htb]
	\centering
	\resizebox{\textwidth}{!}{%
	\begin{tabular}{p{18cm} c p{7cm}}
	\toprule
     Parameter & Units & Description \\
    \midrule
    $U\ts{k} = U\ts{k,ref} + (T\ts{k} - T_{\infty})\pdv{U\ts{k}}{T\ts{k}}\bigg|_{T\ts{k}=T_\infty} \quad \kin{n, p}$ & $\SI{}{\volt}$ & Open circuit potential  \\
    $U\ts{n,ref} = 0.194+1.5\exp(-120.0\frac{c\ts{n}}{c\ts{s,n,max}}) + 0.0351 \tanh((\frac{c\ts{n}}{c\ts{s,n,max}}-0.286)/0.083) - 0.0045\tanh((\frac{c\ts{s,n}}{c\ts{s,n,max}}-0.849)/0.119) - 0.035\tanh((\frac{c\ts{s,n}}{c\ts{s,n,max}}-0.9233)/0.05) - 0.0147\tanh((\frac{c\ts{s,n}}{c\ts{s,n,max}}-0.5)/0.034) - 0.102\tanh((\frac{c\ts{s,n}}{c\ts{s,n,max}}-0.194)/0.142) - 0.022\tanh((\frac{c\ts{s,n}}{c\ts{s,n,max}}-0.9)/0.0164) - 0.011\tanh((\frac{c\ts{s,n}}{c\ts{s,n,max}}-0.124)/0.0226) + 0.0155\tanh((\frac{c\ts{s,n}}{c\ts{s,n,max}}-0.105)/0.029)$ & $\SI{}{\volt}$ & Reference open circuit potential \\
    $U\ts{p,ref} = 2.16216 + 0.07645\tanh(30.834 - 54.4806\frac{c\ts{s,p}}{c\ts{s,p,max}}) + 2.1581\tanh(52.294 - 50.294\frac{c\ts{s,p}}{c\ts{s,p,max}}) - 0.14169\tanh(11.0923 - 19.8543\frac{c\ts{s,p}}{c\ts{s,p,max}}) + 0.2051\tanh(1.4684 - 5.4888\frac{c\ts{s,p}}{c\ts{s,p,max}}) + 0.2531\tanh((-\frac{c\ts{s,p}}{c\ts{s,p,max}} + 0.56478)/0.1316) - 0.02167\tanh((\frac{c\ts{s,p}}{c\ts{s,p,max}} - 0.525)/0.006)$ & $\SI{}{\volt}$ & Reference open circuit potential \\
 $\pdv{U\ts{n}}{T\ts{n}}\bigg|_{T\ts{n}=T_\infty} =
    -1.5(120.0/c\ts{s,n,max})\exp(-120.0\frac{c\ts{s,n}}{c\ts{s,n,max}}) + (0.0351/(0.083c\ts{s,n,max}))((\cosh((\frac{c\ts{s,n}}{c\ts{s,n,max}}-0.286)/0.083))^{-2}) -  (0.0045/(0.119c\ts{s,n,max}))((\cosh((\frac{c\ts{s,n}}{c\ts{s,n,max}}-0.849)/0.119))^{-2}) - (0.035/(0.05c\ts{s,n,max}))((\cosh((\frac{c\ts{s,n}}{c\ts{s,n,max}}-0.9233)/0.05))^{-2}) -  (0.0147/(0.034c\ts{s,n,max}))((\cosh((\frac{c\ts{s,n}}{c\ts{s,n,max}}-0.5)/0.034))^{-2}) - (0.102/(0.142c\ts{s,n,max}))((\cosh((\frac{c\ts{s,n}}{c\ts{s,n,max}}-0.194)/0.142))^{-2}) - (0.022/(0.0164c\ts{s,n,max}))((\cosh((\frac{c\ts{s,n}}{c\ts{s,n,max}}-0.9)/0.0164))^{-2}) - (0.011/(0.0226c\ts{s,n,max}))((\cosh((\frac{c\ts{s,n}}{c\ts{s,n,max}}-0.124)/0.0226))^{-2}) + (0.0155/(0.029c\ts{s,n,max}))((\cosh((\frac{c\ts{s,n}}{c\ts{s,n,max}}-0.105)/0.029))^{-2})$ & $\SI{}{\volt.\kelvin^{-1}}$ & Entropic change \\
    $\pdv{U\ts{p}}{T\ts{p}}\bigg|_{T\ts{p}=T_\infty} =
     0.07645(-54.4806/c\ts{s,p,max})((1.0/\cosh(30.834-54.4806 \frac{c\ts{s,p}}{c\ts{s,p,max}}))^2) + 2.1581 (-50.294/c\ts{s,p,max}) ((\cosh(52.294-50.294 \frac{c\ts{s,p}}{c\ts{s,p,max}}))^{-2}) + 0.14169(19.854/c\ts{s,p,max}) ((\cosh(11.0923-19.8543 \frac{c\ts{s,p}}{c\ts{s,p,max}}))^{-2}) - 0.2051(5.4888/c\ts{s,p,max})((\cosh(1.4684-5.4888 \frac{c\ts{s,p}}{c\ts{s,p,max}}))^{-2}) - (0.2531/0.1316/c\ts{s,p,max}) ((\cosh((-\frac{c\ts{s,p}}{c\ts{s,p,max}}+0.56478)/0.1316))^{-2}) - (0.02167/0.006/c\ts{s,p,max}) ((\cosh((\frac{c\ts{s,p}}{c\ts{s,p,max}}-0.525)/0.006))^{-2})$ & $\SI{}{\volt.\kelvin^{-1}}$ & Entropic change \\
    \midrule
    $D\ts{e} = (5.34 \times 10^{-10}) \times \exp(-0.65 c\ts{e} \times 10^{-3}) \times \exp\left( \frac{E_{D\ts{e}}}{R_g}\left(\frac{1}{T_{\infty}}- \frac{1}{T\ts{k}}\right)\right)$ & $\SI{}{\metre^2.\second^{-1}}$ & Electrolyte diffusivity \\
    $D\ts{s,k} = D\ts{s,k,typ} \quad \kin{n, p}$ & $\SI{}{\metre^2.\second^{-1}}$  & Solid diffusivity \\
    $\kappa\ts{e} = \left(0.0911 + 1.9101 c\ts{e} \times 10^{-3} - 1.052(c\ts{e} \times 10^{-3} )^2 + 0.1554(c\ts{e} \times 10^{-3} )^3\right) \times \exp\left( \frac{E_{\kappa\ts{e}}}{R_g}\left(\frac{1}{T_{\infty}}- \frac{1}{T\ts{k}}\right)\right)$ & $\SI{}{\ohm^{-1}.\metre^{-1}}$ & Electrolyte conductivity \\
    $m\ts{k} = m\ts{k,typ} \exp\left(  \frac{E_{m\ts{k}}}{R_g}\left(\frac{1}{T_{\infty}}- \frac{1}{T\ts{k}}\right)\right) \quad \kin{n, p}$ & $\SI{}{\ampere.\metre^{-2}.(\metre^3.\mol^{-1})^{1.5}}$ & Nominal reaction rate \\
    \bottomrule
	\end{tabular}}
        \caption{Model coefficients taken from Scott Moura's DFN model (based on DUELFOIL parameters). The coefficients are for a graphite negative electrode, an LiPF$_6$ in EC:DMC electrolyte, and a LCO positive electrode.} \label{table:dimensional_coefficients}
\end{table}


\section{Modifications to the SPMe}
\label{app:spme-mods}
The governing equations for the SPMe used here are slightly modified from the version stated in \cite{Marquis2019}. Firstly, we have extended the model to account for thermal effects. Secondly, in \cite{Marquis2019}, a linear diffusion equation for the electrolyte concentration is used by taking the lithium-ion flux in the electrolyte to be of the form
\begin{align}\label{eqn:linear-spme:N_e_k}
    \di{N\ts{e,k}} = -\epsilon\ts{k}^{\text{b}} \di{D\ts{e}}(\di{c\ts{e,$0$}}, \di{T}) \pdv{\di{c\ts{e,k}}}{\di{x}} +     
    \begin{cases}
      \displaystyle 
	  \frac{\di{x} t^+ \di{\mathcal{I}}}{\di{F}\di{L\ts{n}}}, \quad &\text{k}=\text{n}, \\[3mm]
   \displaystyle    \frac{t^+\di{\mathcal{I}}}{\di{F}}, \quad &\text{k}=\text{s}, \\[3mm]
 \displaystyle 	 \frac{(\di{L}-\di{x})t^+ \di{\mathcal{I}}}{\di{F} \di{L\ts{p}}}, \quad &\text{k}=\text{p}.
    \end{cases}
\end{align}
in place of (\ref{eqn:SPMe:N_e_k}). The only difference is the dependence of the diffusion coefficient upon $c\ts{e,0}$ instead of $c\ts{e,k}$. By asymptotically expanding (\ref{eqn:SPMe:N_e_k}) and (\ref{eqn:linear-spme:N_e_k}) in powers of $\mathcal{C}\ts{e}$ (as defined in Table~\ref{tab:conditions}), it can be seen that the two are asymptotically equivalent up to terms of size $\mathcal{O}(\mathcal{C}\ts{e}^2)$. This means that formally the same order of error is introduced by using either version. However, through numerical experimentation, we have found that in some situations, particularly at higher C-rates, utilizing (\ref{eqn:SPMe:N_e_k}) instead of (\ref{eqn:linear-spme:N_e_k}) more accurately recovers the electrolyte concentration and therefore other key variables in the model. 

The other modification is to retain the `$\log$' terms in (\ref{eqn:SPMe:eta_c}), (\ref{eqn:SPMe:phi_e_n}), (\ref{eqn:SPMe:phi_e_s}), and (\ref{eqn:SPMe:phi_e_p}) instead of linearising as in \cite{Marquis2019}. Therefore we convert as follows
\begin{equation}
    \frac{1}{c\ts{e,0}}\left( c\ts{e,$\alpha$} - c\ts{e,$\beta$} \right)\rightarrow \log\left( \frac{c\ts{e,$\alpha$}}{c\ts{e,$\beta$}} \right).
\end{equation}
By making asymptotic expansions of these expressions in terms of powers of $\mathcal{C\ts{e}}$, we can again show that the two forms are asymptotically equivalent up to terms of size $\mathcal{O}(\mathcal{C}\ts{e})$.

\begin{figure}[htbp]
    \centering
    \includegraphics{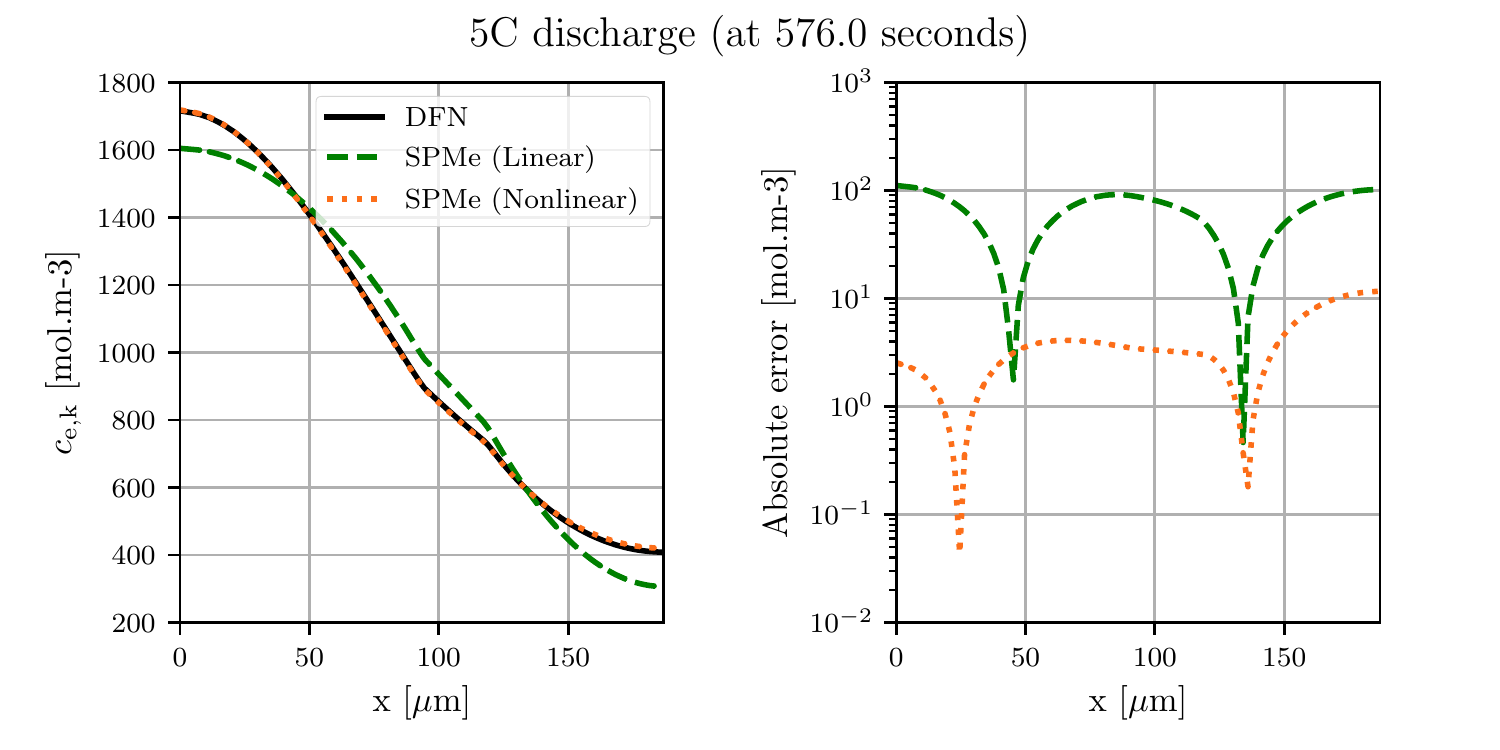}
    \caption{Comparison of electrolyte concentration predicted by the DFN, SPMe (Linear) from \cite{Marquis2019}, and SPMe (Nonlinear) for a \SI{5}{C} discharge using the parameter set in \cite{ecker2015}: (a) electrolyte concentration profile; (b) maximum absolute errors in electrolyte concentration at every time in discharge.}
    \label{fig:electrolyte-comparison}
\end{figure}

In Figure~\ref{fig:electrolyte-comparison}, we present the electrolyte concentration predicted by the DFN, SPMe (linear) from \cite{Marquis2019}, and the SPMe (nonlinear) employed in this paper using the parameters in \cite{ecker2015} and a C-rate of 5. We observe that for this parameter set and C-rate the nonlinear version of the SPMe performs significantly better at recovering the electrolyte concentration. However, if we examine Table~\ref{tab:spme-rmse}, we observe that for lower C-rates both versions of the SPMe recover the terminal voltage to a similar degree of accuracy. Therefore, in some circumstances the slightly simpler version of the SPMe as written \cite{Marquis2019} may be more appropriate.

\begin{table}
    \centering
    \begin{tabular}{c c c c c}
        \toprule
        & \multicolumn{4}{c}{RMSE (mV)} \\
        & 1C & 2.5C & 5C & 7.5C \\
        \midrule
        SPMe (Linear) & 0.6898 & 2.767 & 5.109 & - \\
        SPMe (Nonlinear) & 0.7080&  2.8334&  5.6791&  12.2611 \\
        \bottomrule
    \end{tabular}
    \caption{RMSE for each version SPMe compared to the DFN at different C-rates. Note that the error for SPMe (Linear) at 7.5C is omitted as the model breaks down before the discharge is complete. These results are for the parameter set in \cite{ecker2015}.}
    \label{tab:spme-rmse}
\end{table}

\clearpage

\bibliographystyle{unsrt}
\bibliography{refs}

\clearpage
\section*{Acknowledgements}
This publication is based on work supported by the EPSRC Centre For Doctoral Training in Industrially Focused Mathematical Modelling (EP/L015803/1) in collaboration with Siemens Corporate Technology and BBOXX. This work was also supported with funding provided by The Faraday Institution, grant number EP/S003053/1, FIRG003.

\end{document}